\documentclass[twocolumn,secnumarabic,amssymb, nobibnotes, aps, longbibliography]{revtex4-1}

\usepackage{graphicx}
\usepackage{dcolumn}
\usepackage{bm}
\usepackage{amsmath}

\usepackage{color}
\usepackage{braket}
\usepackage{physics}
\usepackage{chngcntr}
\usepackage{ulem}

\begin{document}

\begin{abstract}
	Dynamic hysteresis effects have been long known to occur in the J-V characteristics of perovskite solar cells (PSCs), with the ionic migration being identified as the primary factor. The hysteretic effects impacted early studies by the uncertainty in the evaluation of the power conversion efficiency, while currently, potential links to degradation mechanisms are in the focus. Therefore, understanding ion migration is a central goal, typically addressed by performing a combined large- and small signal analysis. The reported large capacitive and inductive effects created controversies with respect to the underlying mechanisms, yielding essentially two classes of models, one based on large accumulation capacitances and the other based on ionic modulation of the collected current. We introduce here an equivalent circuit model and interpret these phenomena in terms of recombination current modulation, identifying the distinct contributions from ion current and ionic charge accumulations. These contributions to the recombination current are associated with capacitive and inductive effects, respectively, and we corroborate the numerical simulations with electrochemical impedance spectroscopy (EIS) measurements. These show the role of the recombination currents of photogenerated carriers in producing both capacitive and inductive effects as the illumination is varied. Moreover, we provide a bridging point between the two classes of models and suggest a framework of investigation of defect states based on the observed inductive behavior, which would further aid the mitigation of the degradation effects.
\end{abstract}

\title{Capacitive and inductive effects in perovskite solar cells: the different roles of ionic current and ionic charge accumulation}

\author{Nicolae Filipoiu$^{1,2}$, Amanda Teodora Preda$^{1,2,3}$, Dragos-Victor Anghel$^{1,2,3}$, Roxana Patru$^{4}$, Rachel Elizabeth Brophy$^{5}$, Movaffaq Kateb$^{5}$, Cristina Besleaga$^4$, Andrei Gabriel Tomulescu$^{4}$, Ioana Pintilie$^{4*}$, Andrei Manolescu$^{5}$, George Alexandru Nemnes$^{1,2,3}$}
\email{\vspace{2mm} Corresponding authors: \\
G.A. Nemnes (nemnes@solid.fizica.unibuc.ro), \\
Ioana Pintilie (ioana@infim.ro)} 
\affiliation{$^1$University of Bucharest, Faculty of Physics, 077125 Magurele-Ilfov, Romania}%
\affiliation{$^2$Horia Hulubei National Institute for Physics and Nuclear Engineering, 077126 Magurele-Ilfov, Romania}%
\affiliation{$^3$Research Institute of the University of Bucharest (ICUB), Mihail Kogalniceanu Blvd 36-46, 050107 Bucharest, Romania}%
\affiliation{$^4$National Institute of Materials Physics, Magurele 077125, Ilfov, Romania}%
\affiliation{$^5$Department of Engineering, Reykjavik University, Menntavegur 1, IS-102 Reykjavik, Iceland}%

\maketitle

\section{Introduction}

The perovskite solar cells (PSCs) represent a promising alternative to the well established technologies like silicon-based and thin-film solar cells, due to the significant power conversion efficiency (PCE) \cite{nrel}, obtained at comparatively lower fabrication costs. In addition, silicon-perovskite tandem photovoltaic structures can yield even higher PCEs. Presently, the large number of 2D-3D hybrid perovskites that can be assembled ensures a great versatility of the absorber materials. However, the most problematic aspect which still hinders the commercialization concerns the stability of the PSCs and therefore the possible degradation mechanisms are at the forefront of the current research \cite{doi:10.1021/acs.jpclett.9b00613,10.3389/felec.2021.712785,Guo2021}. These are triggered by the ambient conditions, such as humidity and oxygen, along with the operation conditions of the solar cell. While solar cell encapsulation can address the first two issues, potential detrimental effects due to illumination and applied voltages are required to be addressed by material- and device design. In this respect, detecting and mitigating ion migration is an important goal, as it may provide an early indication on the subsequent stability behavior of the PSC.    

Starting with very early studies, the dynamic hysteretic effects were established as a hallmark in the J-V characteristics of the PSCs \cite{C4EE02465F,doi:10.1021/jz5011187}. The apparent capacitive effects received a lot of attention and various interpretations were proposed, such as ferroelectric domains \cite{doi:10.1021/jz502111u,C4TA04969A,doi:10.1063/1.4890246} of the perovskite layer, a giant dielectric constant \cite{doi:10.1021/jz5011169}, trapping and detrapping of carriers \cite{doi:10.1021/acs.nanolett.7b01211}, large accumulations of charges at the interfaces \cite{doi:10.1021/acs.jpclett.5b02810} and ionic transport \cite{Eames2015,C4EE03664F,C5EE02740C,Meloni2016}. {\it Ab initio} calculations have subsequently dismissed the hypothesis of ferroelectricity as the rotation time scales of the molecular dipoles is in picosecond range \cite{doi:10.1021/acs.jpcc.5b05823,Meloni2016}.

Understanding the dynamic phenomena in PSCs is important from at least two perspectives, which concern a correct evaluation of the solar cell PCE and, possibly, anticipating the solar cell degradation. 
Since the discovery of the hysteretic effects, several models were developed, most employing drift-diffusion equations and equivalent circuits, describing the dynamic J-V characteristics and providing small signal analysis, which we briefly enumerate below.

A drift-diffusion based model reproducing dynamic J-V hysteresis was developed by van Reenen {\it et al.}, which includes ionic migration and trapping, using a finite difference method \cite{doi:10.1021/acs.jpclett.5b01645}. Richardson {\it et al.} also implemented a time-dependent drift diffusion-model, which includes ionic charge buildup, where, in order to cope with the numerical difficulty of the problem, the method of matched asymptotic analysis was employed \cite{C5EE02740C}. Nemnes {\it et al.} introduced a dynamic electrical model (DEM), based on an equivalent circuit which has as key assumption the slow relaxation of the internal device polarization \cite{SolarEnergyMaterialsSolarCells.159.197.2017.Nemnes}. Similarly, Ravishankar {\it et al.} described a surface polarization model (SPM) \cite{doi:10.1021/acs.jpclett.7b00045}, implying large accumulation capacitances. The SPM was extended by Ghahremanirad {\it et al.} to include inductive effects \cite{doi:10.1021/acs.jpclett.7b00415}. The DEM was later reformulated by Anghel {\it et al.} in terms of a non-linear capacitor \cite{Anghel_2019}.
Courtier {\it et al.} derived a simplified surface polarization model, replacing the ion dynamics with non-linear capacitances \cite{courtier2019}, having as starting point the drift-diffusion model described in Ref.\ \cite{C5EE02740C}. The connection between the drift-diffusion simulations and equivalent circuits has been analyzed by Riquelme {\it et al.} \cite{D2CP01338J}, revealing in how far specific electronic and ionic contributions can affect the different equivalent circuit components, which enables their correlation with ion-driven mechanisms. Jacobs {\it et al.} developed a drift-diffusion model, coupling ionic migration to quasi steady-state electron and hole currents, leading to a phase-delayed recombination of photogenerated carriers \cite{doi:10.1063/1.5063259}. Moia {\it et al.} proposed an extensive equivalent circuit model based on an ionically gated transistor behavior \cite{C8EE02362J}.

In addition to the capacitive effects, a peculiar large inductive behavior was evidenced in some PSCs by electrochemical impedance spectroscopy (EIS) technique. So far, the inductive or negative capacitance behavior has been attributed to a kinetic effect induced by ions producing a delay in the surface voltage and charge accumulations \cite{doi:10.1021/acs.jpcc.6b01728,doi:10.1021/acs.jpclett.0c02331}, phase-delayed recombination \cite{doi:10.1063/1.5063259}, electron injection effects \cite{C8EE02362J,Ebadi2019}, ionic mediated recombination \cite{KHAN2021102024} or by action of a chemical inductor \cite{doi:10.1021/jacs.2c00777}, a term used for a generic behavior, which arises from two coupled processes describing a fast-slow dynamics. Based on the chemical inductor phenomenology, the capacitive and inductive effects have been described in halide PSCs and memristors under illumination \cite{10.3389/fenrg.2022.914115}. Inductive effects of chemical origin were observed in the current transients under illumination \cite{doi:10.1021/acsenergylett.2c01252} and were further modeled by a neuron-style model \cite{doi:10.1021/acs.jpcc.2c02729}. A comprehensive review on impedance spectroscopy of perovskite solar cells summarizes the models proposed for explaining the capacitive and inductive effects based on equivalent circuits \cite{doi:10.1021/acs.chemrev.1c00214}.

Here, we introduce an equivalent circuit model which outlines the different roles of the {\it ionic charge current} and {\it ionic charge accumulation} in modulating the recombination current, which stem from the ionic conductor nature of the perovskite layer. Under normal operating conditions, the ions (e.g. iodine ions and/or iodine vacancies) are drifting in the electric field established in the perovskite layer. The ionic current is shown to be responsible for the apparent huge capacitance, while the ionic accumulations produce the large inductive effects. Moreover, in the proposed model, the circuit elements have clear physical interpretations, as they can be associated with the PSC building blocks and processes occurring during solar cell operation.  

The paper is structured as follows. In Section \ref{ecmodels}, the equivalent circuit model is introduced, providing both large- and small signal analysis. This is described by a coupled system of differential equations, which yields the time dependence of the current under illumination and bias conditions in a dynamic regime. Next, in Section \ref{cavscc}, we provide a discussion aimed to bridging the rather different pictures of charge accumulation and charge collection type models found in the literature. Subsequently, the origins of the capacitive and inductive effects are discussed in Section \ref{simexp} as part of the small signal analysis. Here, the two different mechanisms for recombination current dephasing are outlined. The observed trends and features are supported by experimental EIS measurements performed on PSCs with standard configuration, FTO/c-TiO$_2$/m-TiO$_2$/CH$_3$NH$_3$PbI$_3$/Spiro-OMeTAD. Further details concerning numerical simulations are found in the Appendix Supplemental Material (SM).

\section{Equivalent circuit models: large- and small-signal analysis}
\label{ecmodels}

\subsection{Large-signal modeling}

\begin{figure}[t]%
\begin{center}
  \includegraphics[width=0.98\linewidth]{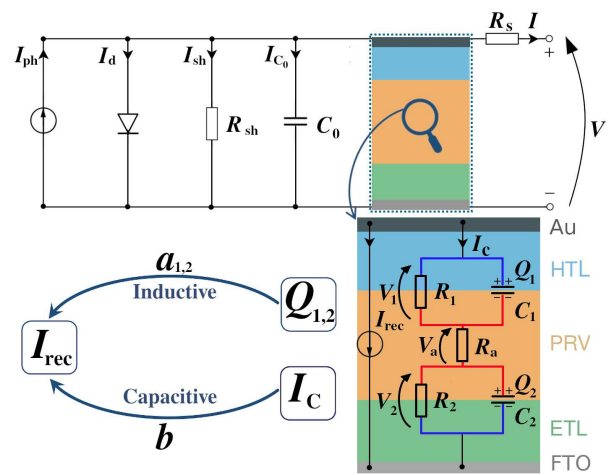}
\end{center}
	\caption{Equivalent circuit describing capacitive and inductive effects. The standard elements like the current source describing the photogenerated current ($I_{\rm ph}$), diode element, the shunt and series resistances and the geometrical capacitance reproduce a typical almost hysteresis-free I-V characteristics. The circuit block includes the ionic capacitors ($C_{1,2}$), ionic absorber resistance ($R_{\rm a}$) and ionic capacitors' loss resistances ($R_{1,2}$) and it controls the recombination current ($I_{\rm rec}$). The circuit elements are overlaid on the structural components of the PSC, in the standard configuration, with FTO and gold electrodes. The capacitive and inductive effects are linked to the ionic current ($I_{\rm c}$) and ionic charge accumulations ($Q_{1,2}$), respectively. The pathways of ions and of electrons/holes are depicted in red and blue colors, respectively.}
    \label{EQC_mod_3}
\end{figure}

We introduce the equivalent circuit model depicted in Fig.\ \ref{EQC_mod_3}, which consistently explains the typical hysteretic effects evidenced in the dynamic J-V characteristics: the normal hysteresis (NH) and inverted hysteresis (IH) behavior, the current bump in the reverse characteristics for positive bias pre-poling and the bias scan rate dependence of the hysteresis magnitude. It features the standard elements, like the current source corresponding to the photogenerated current $I_{\rm ph}$, the diode element introducing a recombination current $I_{\rm d}$, the shunt resistance $R_{\rm sh}$, accounting for a recombination current $I_{\rm sh}$ and the series resistance $R_{\rm s}$. We may also add to this group the relatively small geometrical capacitance $C_0$. These five standard elements ($I_{\rm ph}$, $I_{\rm d}$, $R_{\rm sh}$, $R_{\rm s}$, $C_0$) represent the back-bone of our circuit model. However, the hysteretic effects are reproduced by including two essential elements, namely the $R$-$C$ circuit block and the additional current source corresponding to the ionic modulated recombination current, $I_{\bf rec}$. The $R$-$C$ block includes the two parallel groups, $R_1$-$C_1$ and $R_2$-$C_2$, in series with the perovskite (PRV) absorber ionic resistance $R_{\rm a}$. In our model, which belongs to the charge collection type models, the capacitors $C_1$ and $C_2$, with instantaneous applied voltages $V_1(t)$ and $V_2(t)$, describe the ionic charge accumulation at the interfaces between the absorber and the hole transporter layer (HTL) and the electron transporter layer (ETL), respectively. The corresponding charges are denoted by $Q_{1}$ and $Q_{2}$. The resistance elements $R_1$ and $R_2$ account for the electron/hole-ion charge neutralization at the two interfaces and can be regarded as loss resistances for the ionic capacitors. In this picture, the current $I_{\rm c}$ denotes the ionic current flowing inside the perovskite absorber layer. Later, we shall discuss by comparison the other perspective, which assumes large photogenerated charge accumulations as a basic mechanism for the hysteretic effects. In a typical setup of a dynamic measurement, the applied voltage $V(t)$ is varied with a constant scan rate $\alpha=dV/dt$, which results in a collected current $I(t)$ in the external circuit.

In accordance with Fig.\ \ref{EQC_mod_3}, the equivalent circuit model is described by the system of coupled equations (\ref{V1dt}--\ref{ieq}):
\begin{equation}
\label{V1dt}
	\frac{\partial V_1}{\partial t} = -\frac{1}{C_1}\left(\frac{R_{\rm a}+R_1}{R_{\rm a}R_1}\right)V_1 + \frac{V+IR_{\rm s}-V_2}{R_{\rm a}C_1},
\end{equation}
\begin{equation}
\label{V2dt}
	\frac{\partial V_2}{\partial t} = -\frac{1}{C_2}\left(\frac{R_{\rm a}+R_2}{R_{\rm a}R_2}\right)V_2 + \frac{V+IR_{\rm s}-V_1}{R_{\rm a}C_2},
\end{equation}
\begin{equation}
\label{ieq}
	I_{\rm ph} - I_{\rm d} - I_{\rm sh} - I_{\rm c0} - I_{\rm c} - I_{\rm rec} - I = 0 ,
\end{equation}	
which provides the time dependence of the state variables $V_1$, $V_2$ and $I$, having as input the applied voltage $V(t)$ and the initial conditions, $V_1(t=0)$, $V_2(t=0)$ and $I(t=0)$. These are usually established by applying a pre-poling voltage $V_{\rm pol}$ long enough to reach stationarity. Subsequently, a continuous reverse-forward scan is performed, i.e. the voltage $V$ is varied from an initial voltage $V_0$ to short-circuit and back. The time-dependent solution of the system (\ref{V1dt}--\ref{ieq}) ultimately yields the dynamic $I(V)$ characteristics.

Equation (\ref{ieq}) contains different current contributions, which are functions of the state variables and are specified in the following: the diode current,
$I_{\rm d} = I_{\rm s} \left[\exp\left(\frac{q_{\rm e}(V+IR_{\rm s})}{n_{\rm id}k_{\rm B}T}\right)-1\right]$, which stands for the effective junction recombination in the absence of ions, where
$I_{\rm s}$ is the diode saturation current and $n_{\rm id}$ is the ideality factor; the shunt recombination current, $I_{\rm sh} = (V+IR_{\rm s})/R_{\rm sh}$, due to junction short-circuits which may occur due to pinholes defects and metal diffusion, being rather small for a performant PSC; the current due to the geometrical capacitance, $I_{\rm c0} = C_0\left(\frac{\partial V}{d t}+R_{\rm s}\frac{\partial I}{\partial t}\right)$; the current due to ionic capacitance, $I_{\rm c} = \frac{V+IR_{\rm s}-V_1-V_2}{R_{\rm a}}$.

Our central assumption is that the recombination current $I_{\rm rec}$, which brings the essential contribution to the hysteretic effects in the J-V characteristics, as well as to the capacitive and inductive effects, is modulated by both ion charge accumulation and ionic current:
\begin{equation}
\label{irec}
	I_{\rm rec} = I_{\rm rec0} + \sum_{i=1}^2 a_i Q_{i} + b I_{\rm c},
\end{equation}	
where the first term, $I_{\rm rec0}$, is the reference recombination current, in the absence of the ionic influence, which simply regauges the photogenerated current, while $a_i$ and $b$ parameterize the inductive and capacitive recombination current contributions, $I_{\rm rec}^{\rm L} = \sum_{i=1}^2 a_i Q_{i}$ and $I_{\rm rec}^{\rm C} = b I_{\rm c}$, respectively. These contributions are illustrated in detail in Fig.\ S1 in the SM. The control  exerted on the recombination current by the ionic charge accumulations ($Q_{1,2}$) and by the electric field in the perovskite, ${\mathcal E} \sim I_{\rm c}$, is modeled as a current source in Fig.\ \ref{EQC_mod_3}.

Note that all the currents indicated in Eq.\ (\ref{ieq}) may be replaced by current densities, e.g. $J=I/{\mathcal A}$, where ${\mathcal A}$ is the active area of the solar cell.

The observed capacitances are deemed to have an exponential voltage dependence beyond $V_{\rm oc}$, as  noted in Ref.\ \cite{SolarEnergyMaterialsSolarCells.159.197.2017.Nemnes} by the sharp increase of the initial polarization under pre-poling conditions, later reformulated as a $C(V)$ dependence \cite{Anghel_2019} and revealed experimentally by C-V measurements \cite{doi:10.1021/acs.jpcc.8b03948,RAVISHANKAR2018788}. Here, we assume the ionic interface capacitances $C_i$ ($i=1,2$) are functions of voltage, given by:
\begin{equation}
\label{C_V}	
	C_i(V_{i}) = \bar{C}_{0i} + \bar{C}_{1i} \exp\left(\frac{q_{\rm e}V_{i}}{n_c k_{\rm B}T}\right), 
\end{equation}	
where $V_{i}$ are instantaneous voltages on the two capacitors and $\bar{C}_{0i}$, $\bar{C}_{1i}$, $n_c$ parametrize the exponential behavior \cite{Anghel_2019}, while $q_{\rm e}$ and $k_{\rm B}$ are the electron charge and Boltzmann constant, respectively. Hence, the ionic charges are given by:
\begin{equation}
	Q_{i}(V_{i}) = \bar{C}_{0i}V_{i} + \bar{C}_{1i} \frac{n_c k_{\rm B}T}{q_{\rm e}}
	 		\left[ \exp\left(\frac{q_{\rm e}V_{i}}{n_c k_{\rm B}T}\right) - 1 \right] .
\end{equation}
The rather small geometrical capacitance $C_0\sim0.1$ $\mu$F has a negligible effect on the dynamic J-V characteristics. In contrast, the ionic capacitances $C_{1,2}\sim10 - 100$ $\mu$F are a 2-3 orders of magnitude larger, but still below the huge apparent capacitances observed in the impedance spectroscopy measurements, $C_{\rm app}\sim10 - 100$ mF.

\begin{figure}[t]%
\begin{flushleft}
	(a) $V_{\rm pol}>V_{\rm oc}$ $\rightarrow$ high collection rate   	
  \includegraphics[width=0.98\linewidth]{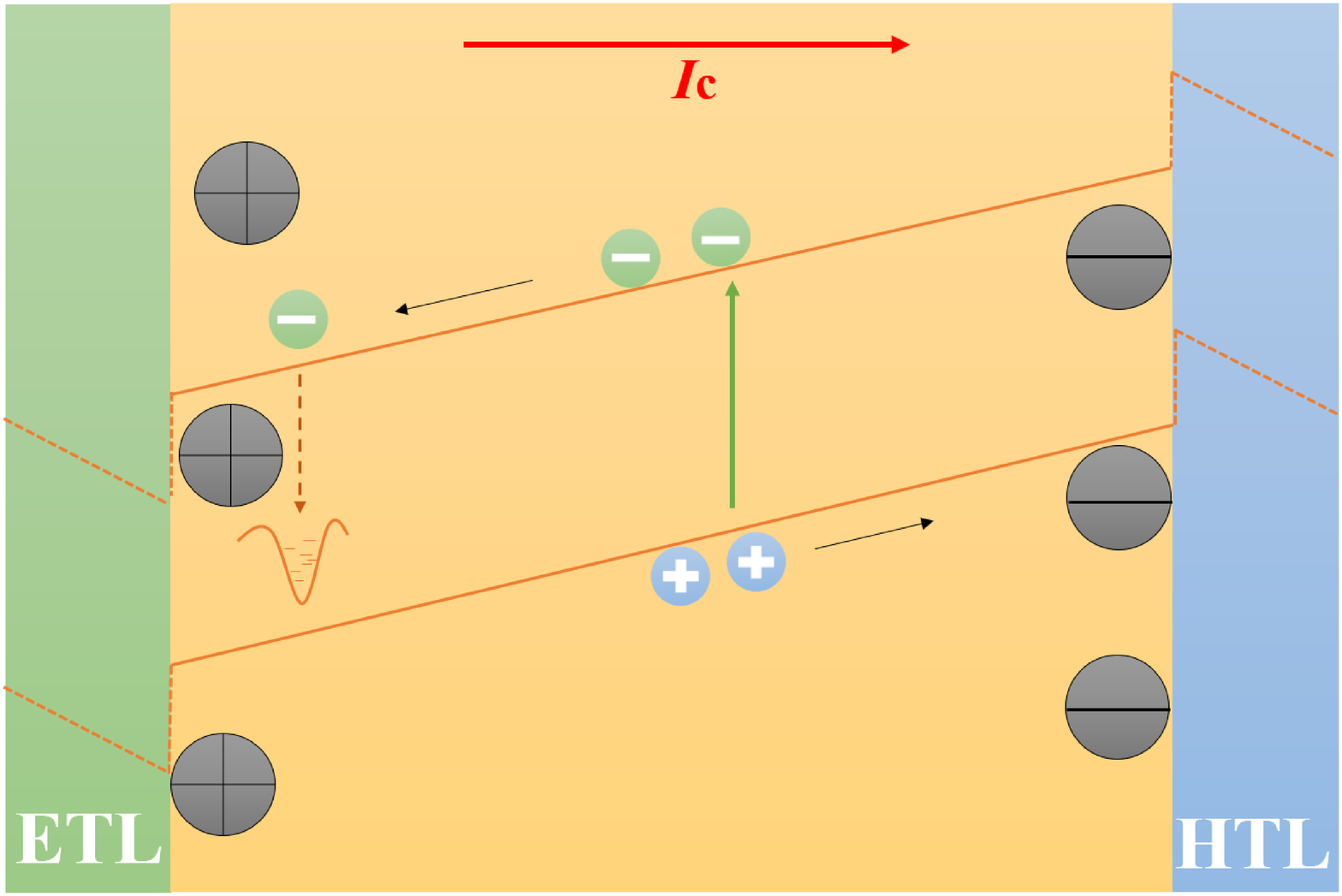}\\
(b) $V_{\rm pol}<0$ $\rightarrow$ high recombination   	
  \includegraphics[width=0.98\linewidth]{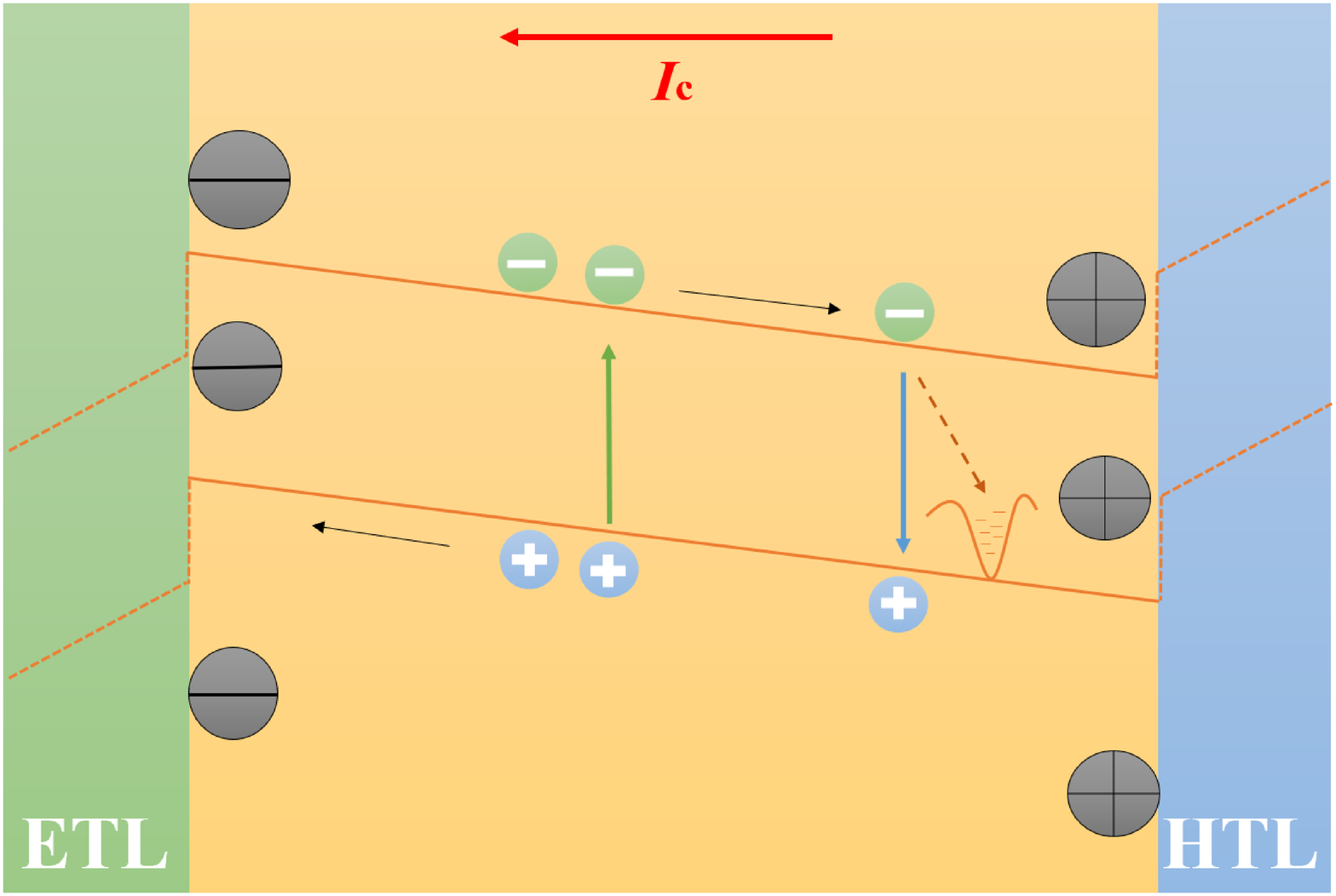}\\
  \includegraphics[width=0.98\linewidth]{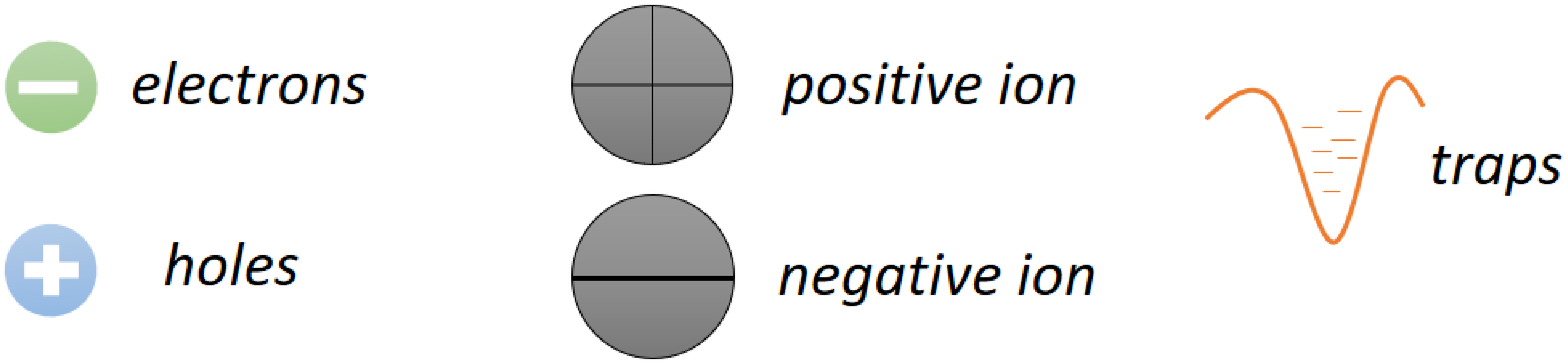}
\end{flushleft}
	\caption{Recombination mechanisms responsible for the capacitive and inductive effects, triggered by ionic accumulations, respectively: instantaneous electric field and ionic-induced trap defects. Two working conditions are illustrated: snapshots during a reverse scan ($dV/dt<0$), in the bias range $0<V<V_{\rm oc}$, after prepoling with (a) $V_{\rm pol}>V_{\rm oc}$ and (b) $V_{\rm pol}<0$. A favorable electric field ${\mathcal E}$ enhances charge collection, reducing recombination, leading to the appearance of the current bump. In contrast, large band-to-band recombination occurs for negative pre-poling bias. In both cases, trap-assisted recombination at the interfaces is associated with inductive effects evidenced in EIS experiments.  
	}
    \label{rec}
\end{figure}

Equation (\ref{irec}) introduces two different recombination current contributions, which correspond to ionic charge accumulations and ionic currents. The second term corresponds to an electron-hole recombination current which is triggered by ion accumulation at the interfaces. High local densities of defects produced by ion accumulations $Q_i$ induce proportionally large recombination of the photogenerated carriers, weighted by the coefficients $a_i$, and shall be correlated with the inductive effects. The third term represents the recombination current contribution due to the electric field in the bulk of the perovskite, ${\mathcal E}$, which is indeed related to the ion current, as stated by Ohm's law $J_{\rm c} = \sigma_{\rm ion} {\mathcal E}$, where $\sigma_{\rm ion}$ is ionic conductivity of the absorber. The parameter $b$ rescales the ionic current $ I_{\rm c}$ and, therefore, the resulting recombination current component is driven in phase with the ionic current, yielding the apparent huge capacitive effects. 
Although the parameters $I_{\rm rec0}$, $a_i$ and $b$ may have, in general, a complex dependence on the illumination and bias conditions, we will show that many of the experimentally observed features can be reproduced with a rather simple assumption: linear dependence on the photogenerated current. Specifically, we shall consider $I_{\rm rec0} = \lambda_0 I_{\rm ph}$, $a_i = \lambda_{a_{0i}} + \lambda_{a_i} I_{\rm ph}$ and $b = \lambda_b I_{\rm ph}$. For the dark condition, $I_{\rm rec0}=0$ and $b$ vanishes, leaving the small ionic capacitance contributions from $C_{1,2}$, while there is still a small inductive contribution given by $ \lambda_{a_{0i}} \times Q_i$, as typically found in the EIS experiments. In the subsequent analysis of inductive effects, we consider positive applied voltages ($V>0$) so that $Q_i>0$, while for negative voltages ($V<0$) the absolute values of ionic charges, $|Q_i|$ should be used, since the recombination current depends on the number of ions and not strictly on their sign, although the $(\lambda_{a_{01}},\lambda_{a_i})$ coefficients would depend on the ionic species like iodine($-$), methylammonium($+$) and their vacancies. The functional dependence of $I_{\rm rec}$ in Eq.\ (\ref{irec}) assumes relatively low recombination rates compared to $I_{\rm}$ and it may be further amended.

The competition between the two types of recombination currents, represented by the 2$^{nd}$ and 3$^{rd}$ terms in Eq.\ (\ref{irec}), gives the overall capacitive or inductive behavior, as illustrated in Fig.\ \ref{rec}. Here, two initial conditions are considered -- pre-poling at large positive and negative biases, while the PSC is operated in the reverse scan ($dV/dt<0$) starting from the open circuit voltage. The band structures in the active region are depicted in Fig.  \ref{rec} (a) and (b), respectively, and it shows the action of the electric field produced by the ionic charges accumulated at the interfaces, following pre-poling. While for $V_{\rm pol}>V_{\rm oc}$ the electric field enhances the charge collection, the pre-poling condition $V_{\rm pol}<0$ followed by the same reverse scan produces the opposite situation: the electric field hinders the charge collection and the band-to-band recombination is increased. It follows that the orientation of the electric field controls the recombination current {\it in phase} with the ionic current and therefore, under illumination, it is perceived as a large capacitance. Moreover, the accumulated ionic charges, particularly at the interfaces, induce Shockley-Reed-Hall type recombinations, leading to inductive effects. This was recently conjectured by Khan {\it et al.} from EIS and space charge-limited current measurements \cite{KHAN2021102024}.

Naturally, the Ansatz proposed in Eq. (\ref{irec}) can be further refined to accommodate more specific experimental conditions. However, it provides a minimal assumption which can reproduce both capacitive and inductive effects, as pointed out by the subsequent small-signal analysis.

\subsection{Small-signal modeling}

Impedance spectroscopy is a non-invasive electrical investigation technique, which can provide valuable information regarding the internal processes occurring during the operation of the PSCs. Starting with the large-signal equivalent circuit shown in Fig.\ \ref{EQC_mod_3}, governed by Eqs.\ (\ref{V1dt}-\ref{ieq}), one obtains the small-signal circuit, by excluding the constant current source element ($I_{\rm ph}$) and replacing the diode element with dynamic resistance $R_{\rm d} = n k_{\rm B} T / (q_{\rm e} I_{\rm d})$ [see Fig.\ S5 and Table I in the SM]. Applying a small-signal voltage $v(t)$ with angular frequency $\omega$, at a given working point set by the external voltage $V = V_{\rm wp}$, which corresponds to a tuple $(V_1,V_2,I)$, one obtains the small current variations $i(t)$ from the current balance, similar to Eq.\ (\ref{ieq}):
\begin{equation}
	-i = i_{\rm d} + i_{\rm sh} + i_{C_0} + i_C + i_{\rm rec},	
\end{equation}
where we may lump together the three standard parallel impedances, corresponding to the diode element, shunt resistance and geometrical capacitor into 
$Z_0 \equiv R_{\rm d} \parallel R_{\rm sh} \parallel Z_{C_0}$, where $Z_{C_0} =1/(j\omega C_0)$ is the complex impedance associated with $C_0$:
\begin{equation}
Z_0 = \left[\frac{1}{R_{\rm d}} + \frac{1}{R_{\rm sh}} + \frac{1}{Z_{C_0}} \right]^{-1} . 
\end{equation}

Using Eq.\ (\ref{irec}) we may write the recombination current around the working point as:
\begin{equation}
	I_{\rm rec}(V) + i_{\rm rec} = I_{\rm rec0} + \sum_{i=1}^2 a_i Q_i(V_i+v_i) + b I_{\rm C}(V+v),
\end{equation}	
which yields: 
\begin{equation}
\label{sm-irec}	
	i_{\rm rec} = \left[ \sum_{i=1}^2 a_i C_i(V_{i}) Z_{R_i C_i} + b \right] \times i_C,
\end{equation}	
where $Z_{R_i C_i} = R_i/(j\omega R_i C_i + 1)$ are the equivalent impedances of the two R-C groups and $v_i = Z_{R_i C_i} i_C$ are the corresponding small-signal voltages. In Eq.\ (\ref{sm-irec}), $i_{\rm rec}$ has two components: the inductive component, $i_{\rm rec}^{\rm L} = \sum_{i=1}^2 a_i C_i(V_{i}) Z_{R_i C_i} \times i_C$, and the capacitive component, $i_{\rm rec}^{\rm C} = b \times i_C$. 

Taking into account that the small-signal ionic current $i_C = (v+iR_{\rm s})/Z_C$, where $Z_C = Z_{R_1 C_1} + R_{\rm a} + Z_{R_2 C_2}$ is the total impedance of the ionic circuit block, we can express the equivalent impedance of the small-signal circuit as:
\begin{equation}
Z = \frac{v}{-i} = R_{\rm s} + 
	\left[\frac{1}{Z_0} + \sum_{i=1}^{2} a_i C_i(V_{i}) \frac{Z_{R_i C_i}}{Z_C} + \frac{(b+1)}{Z_C} \right]^{-1} .
\end{equation}
The impedance function includes $R_{\rm s}$ in series with the parallel group formed by $Z_0$, $Z^{\rm eq}_{\rm L}$ and $Z^{\rm eq}_{\rm C}$, where the equivalent impedances, reflecting inductive and capacitive components, are given by:
\begin{eqnarray}
\label{ZL}
Z^{\rm eq}_{\rm L} &=&   \left[ \sum_{i=1}^{2} a_i C_i(V_{i}) \frac{Z_{R_i C_i}}{Z_C} \right]^{-1} , \\
\label{ZC}
Z^{\rm eq}_{\rm C} &=&   \frac{Z_C}{(b+1)} .
\end{eqnarray}	

A simplified one-capacitor model shown in Fig.\ S5(b) in the SM can be obtained from the two-capacitor model by setting $R_1\rightarrow\infty$, $R_2\rightarrow0$. In this case, denoting $C_1=C_{\rm ion}$, $R_a=R_{\rm ion}$, $a=a_1$, Eqs.\ (\ref{sm-irec}-\ref{ZC}) are simplified to:
\begin{eqnarray}
\label{sm-irec-1cap}
	i_{\rm rec} &=& -j \frac{a}{\omega} \times i_C + b \times i_C, \\
\label{sm-irec-Z}
	Z = \frac{v}{-i} &=& R_{\rm s} + 
	\left[\frac{1}{Z_0} + \frac{1}{Z_{\rm RL}} + \frac{1}{Z_{\rm RC}} \right]^{-1}, \\        
\label{Z1L}
	Z_{\rm RL} &=& \frac{1}{a C_{\rm ion}} + j\; \omega \frac{R_{\rm ion}}{a}, \\
\label{Z1C}
	Z_{\rm RC} &=&   \frac{R_{\rm ion}}{(b+1)} - j \frac{1}{\omega C_{\rm ion} (b+1)} .
\end{eqnarray}	
In Eq.\ (\ref{sm-irec-1cap}) $i_{\rm rec}^{\rm L} = -j \frac{a}{\omega} \times i_C$ and $i_{\rm rec}^{\rm C} =  b \times i_C$, while from Eqs.\ (\ref{Z1L}) and (\ref{Z1C}) we can identify the resistive components $R_{\rm L} = 1/(a C_{\rm ion})$, $R_{\rm acc} = R_{\rm ion}/(b+1)$, the inductance $L = R_{\rm ion}/a$ and the capacitance $C_{\rm acc} = (b+1) C_{\rm ion}$. These quantities are further discussed in Section \ref{cavscc}.

Furthermore, one can define the apparent capacitance, which is typically measured by EIS, as:
\begin{equation}
	C_{\rm app}(\omega) = \omega^{-1} \mbox{Im}(Z^{-1}).
\end{equation}

\section{Charge accumulation vs. charge collection models}
\label{cavscc}

\begin{figure}[t]%
\begin{flushleft}
(a) Charge accumulation (CA) model
  \includegraphics[width=0.95\linewidth]{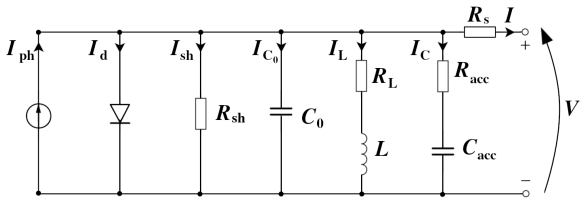}\\
(b) Charge collection (CC) model 	
  \includegraphics[width=0.95\linewidth]{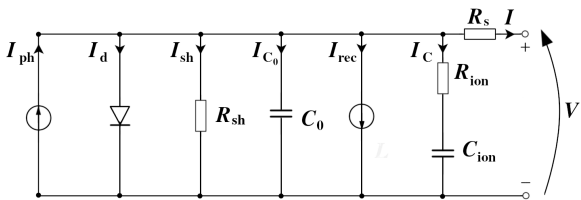}
\end{flushleft}
	\caption{Bridging CA and CC models: (a) Charge accumulation model with inductive component ($L$). (b) Charge collection model based on ion modulated recombination, {\it without} an explicit inductive element. The one-capacitor model is obtained from the two-capacitor model by setting $R_1\rightarrow\infty$, $R_2\rightarrow0$, with $C_1=C_{\rm ion}$ and $R_a=R_{\rm ion}$. Although CA and CC models correspond to rather different microscopic descriptions, they yield the same formal behavior provided the time scales involved are the same, $\tau = R_{\rm acc}C_{\rm acc} = R_{\rm ion}C_{\rm ion}$.}
    \label{CA-1cap}
\end{figure}

There has been a lot of debate concerning the physical origins of the dynamic J-V hysteresis. The very large capacitances \cite{doi:10.1021/acs.jpclett.5b02273,doi:10.1021/acs.jpclett.5b02810,doi:10.1021/acs.jpcc.8b03948,RAVISHANKAR2018788} observed in typical impedance spectroscopy measurements hinted towards possible large accumulations of ions and photogenerated carriers at the interfaces. Based on this assumption, one class of models, here termed as {\it charge accumulation (CA) models}, provided a rather accurate description of the hysteretic effects. In this class we may include the rather generic equivalent circuit models proposed by Seki \cite{doi:10.1063/1.4959247} and Nemnes {\it el.} \cite{SolarEnergyMaterialsSolarCells.159.197.2017.Nemnes}, a surface polarization model accounting for large ionic and electron accumulation charges by Ravishankar {\it al.} \cite{doi:10.1021/acs.jpclett.7b00045}, with additional inductive elements \cite{doi:10.1021/acs.jpclett.7b00415} and drift-diffusion type models of Lopez-Varo {\it et al.} pointing out a similar phenomenology \cite{https://doi.org/10.1002/aenm.201702772}. On the other hand, Tress {\it et al.} proposed a mechanism based on the recombination and charge collection influenced by the electric field in the bulk of the perovskite \cite{C4EE03664F}. This hypothesis defines another class of models, which rely on an ionic modulated recombination current either by the internal electric field in the perovskite or by changing extraction barriers, resulting in a modified {\it charge collection} (CC) efficiency. Subsequently, along these main lines, the ionic contribution to the hysteretic effects was illustrated by the models of Jacobs {\it et al.} \cite{doi:10.1063/1.5063259} and Moia {\it et al.} \cite{C8EE02362J}. The microscopic picture of CC models is alternative to the CA models, as they do not require a large charge accumulation, but rather a small ionic charge influences the charge collection properties. 

In the following we provide a bridging point between a typical CA and our CC model, showing that both can reproduce the typical features of the dynamic hysteresis. To this end, we simplify the two-interface model introduced in Fig.\ \ref{EQC_mod_3} and consider only one interface, i.e. one capacitor in series with a resistor, as it already provides the core physics of the device operation. The CA model considered here is depicted in Fig. \ref{CA-1cap}(a), being similar to the equivalent circuit model with inductive elements from Ref. \cite{doi:10.1021/acs.jpclett.0c02459}, while the corresponding CC model is shown in Fig.\ \ref{CA-1cap}(b).

The hallmark of the hysteretic effects is the relatively large time scale $\tau$, in the range of seconds to a few hundreds of seconds. From the point of view of an R-C circuit, a large time scale $\tau=RC$ can be reproduced either by a large capacitance or a large resistance. The physical interpretation and magnitude of these two elements is quite different in the two types of models. In the case of the CA model, the accumulation capacitance, which includes ions and photogenerated carriers is assumed to be large $C_{\rm acc} \sim 0.1$ F, while the resistance is rather small, characteristic to electron/hole conduction, $R_{\rm acc} \sim 100$ $\Omega$. In contrast, within the CC model, the same two quantities are interpreted as a much smaller ionic capacitance $C_{\rm ion} \sim 100$ $\mu$F and a correspondingly larger ionic resistance $R_{\rm ion} = 10^5$ $\Omega$. However, both models can be calibrated to yield the same time scale $\tau = R_{\rm acc}C_{\rm acc} = R_{\rm ion}C_{\rm ion}$, becoming formally equivalent. 

A recent experimental paper reported an ionic
conductivity of $7.7\times 10^{-9}$~S/cm in the MAPI (CH$_3$NH$_3$PbI$_3$)
perovskite \cite{Zhao2021}, i.\ e. a resistivity $\rho=1.3 \times 10^8$
$\Omega$cm. Considering a perovskite layer of 300 nm thickness ($\ell$)
with an exposed area (${\mathcal A}$) of about 0.1 cm$^2$  (Section \ref{expval}) our
ionic resistance $R_{\rm ion}=\rho \ell /{\mathcal A} = 10^5$ $\Omega$ corresponds to
$\rho \approx 3 \times 10^8$  $\Omega$cm, close to the experimental value.
According to molecular dynamics simulations, in MAPI, the mobility of
the iodine vacancies is $\mu=1.7\times 10^{-4}$ cm$^2$/Vs, and it is
higher than of the interstitial iodide, $\mu=2.8\times 10^{-4}$ cm$^2$/Vs
\cite{Delugas2016}. The ionic conductivity $\sigma=q_e n \mu$ also depends
on the concentration of the ionic charge carriers $n$.  Assuming that the
dominant contribution to the ionic current is the migration of vacancies,
our choice of the resistivity corresponds to 10$^{14}$ vacancies/cm$^3$.
This is a small fraction of the total expected vacancy concentration,
estimated between $10^{17}-10^{20}$ cm$^{-3}$ \cite{Walsh2015}. But, in
reality, a large number of vacancies are trapped at the grain boundaries,
where the total number of trap states is between $10^{16}-10^{18}$
cm$^{-3}$ \cite{Draguta2016}, or at the interfaces between MAPI and the
transporter layers as indicated in Fig.~(\ref{rec}). So, it is reasonable to 
assume that only a small amount of ionic charges contribute to the ionic current.

Taking into account the migration of a single ion species, accumulating at one of the two interfaces, Eq.\ (\ref{irec}) simplifies to:
\begin{equation}
\label{irec1cap}
	I_{\rm rec} = I_{\rm rec0} + a Q_{\rm c} + b I_{\rm c}
		    = I_{\rm rec0} + a Q_{\rm c} + b \frac{\partial Q_{\rm c}}{\partial t} .
\end{equation}	
In a first step, by ignoring the inductance branch in the CA model in Fig.\ \ref{CA-1cap}(a) which corresponds to $a=0$ in the CC model in Eq.\ (\ref{irec1cap}), the two models become formally equivalent. In a typical reverse-forward scan the relatively large $I_{\rm c}^{\rm CA}$ current obtained in the CA model, corresponding to the discharging and recharging of the $C_{\rm acc}$ capacitor, is precisely matched in the CC model by the $b \times I_{\rm c}^{\rm CC}$ contribution to the recombination current $I_{\rm rec}$. The net photogenerated currents are the same, i.e. $I_{\rm ph}^{\rm CA} = I_{\rm ph}^{\rm CC} - I_{\rm rec0}$. Therefore, the ionic current $I_{\rm c}^{\rm CC}$ is rather small, but it controls the recombination current, which is significantly larger, by a factor $b \sim 10^3$. Consequently, in the small signal analysis, the impedance of the $R_{\rm acc}-C_{\rm acc}$ branch is equivalent to the impedance obtained by the $R_{\rm ion}-C_{\rm ion}$ branch together with the modulated current source $I_{\rm rec}$: $Z_{\rm C}^{\rm CC} = Z_{\rm C}^{\rm CA}/(b+1)$. Furthermore, the recombination current component $a Q_{\rm c}$ of the CC model corresponds to the $R_{\rm L}$-$L$ branch in the CA model. This shows that, technically, a black-box approach of the type of the CA model reproduces the capacitive and inductive effects, as described by the CC model, which is based on the ionic modulated recombination and is physically more meaningful.   

In order to account for the inductive (negative capacitance) effects, many equivalent circuit descriptions include explicit inductive elements \cite{doi:10.1021/acs.chemrev.1c00214}. However, although the EIS data can be accurately reproduced, our approach gives a direct physical interpretation of the inductive effects, which relies also on an ionic modulated recombination current, but this time on the ionic charge accumulation ($a\times Q_{\rm c}$) instead of the ionic current ($b\times \frac{\partial Q_{\rm c}}{\partial t}$), the latter being related to the bulk electric field in the absorber. In this context, Eq.\ (\ref{irec1cap}) provides a unified description of both capacitive and inductive effects by means of modulated recombination currents, at the same time outlining the specific physical mechanism for each behavior.  

The advantage of the CA type models relies on the use of standard circuit elements (capacitors and inductive elements), in contrast to capacitor-voltage modulated current-sources present in the CC type models.  However, the huge values obtained for the capacitances and inductances in the CA type models require a careful interpretation of the capacitive/inductive currents in terms of recombination currents. On the other hand, CC type models have a direct physical interpretation, allowing a connection between the circuit elements and the building blocks and processes within the PSCs. 

\section{Simulations and experimental validation}
\label{simexp}

In the following sections we perform both large- and small signal analysis, focusing on the main features that are observed in typical experiments. The large signal analysis comprises J-V simulations performed with different voltage poling and illumination conditions, while the influence of the scan rate and characteristic time scales on the hysteresis magnitude is shown. On the other hand, the small signal analysis reveals the capacitive and inductive effects, which are here identified to follow from two distinct recombination processes, namely the action the bulk electric field and ionic-induced defect recombination, respectively. These features are matched by experimental data.

We consider a set of equivalent circuit parameters which we assign to the reference PSC. In the subsequent discussions some of them shall be modified in order to point out different behaviors concerning capacitive and inductive effects. The following circuit model parameters are used in the reference PSC configuration: series resistance $R_{\rm s} = 50$ $\Omega$; shunt resistance $R_{\rm sh} = 6$ k$\Omega$; photogenerated current $I_{\rm ph} = 2$ mA; diode parameters $I_{\rm s} = 10^{-13}$ A and $n_{\rm id} k_{\rm B}T = 40 $ meV; geometrical capacitance $C_0 = 0.079$ $\mu$F; PSC area ${\mathcal A} = 0.09$ cm$^{-2}$; absorber (ionic) resistance $R_{\rm a} = 500$ k$\Omega$; ionic capacitor parametrization in Eq.\ (\ref{C_V}) (one capacitor case) $\bar{C}_{01} = 10$ $\mu$F, $\bar{C}_{11} = 10^{-14}$ $\mu$F, $n_c k_{\rm B}T = 26$ meV. In this initial PSC configuration only the capacitive effects are accounted for ($a_i=0$, $b=1000$), while the ionic capacitor losses are initially neglected ($R_{1,2}\rightarrow\infty$). The parametrization of $I_{\rm rec}$ as a function of $I_{\rm ph}$ is given by: $\lambda_0=0.5$, $\lambda_{a_{0i}}=10$ $s^{-1}$, $\lambda_{a_{i}}=50$ mC$^{-1}$, $\lambda_b = 500$ mA$^{-1}$. A typical scan rate $\alpha = 20$ mV/s is used in the subsequent large signal analysis. This set of parameters reproduces the experimentally observed hysteretic effects \cite{SolarEnergyMaterialsSolarCells.159.197.2017.Nemnes,doi:10.1021/acs.jpcc.7b04248,NEMNES2018976} and also the main trends in the EIS measurements, as it shown in Section \ref{expval}.

\subsection{Large-signal analysis}

\begin{figure}[t]%
\begin{center}
  \includegraphics[width=0.9\linewidth]{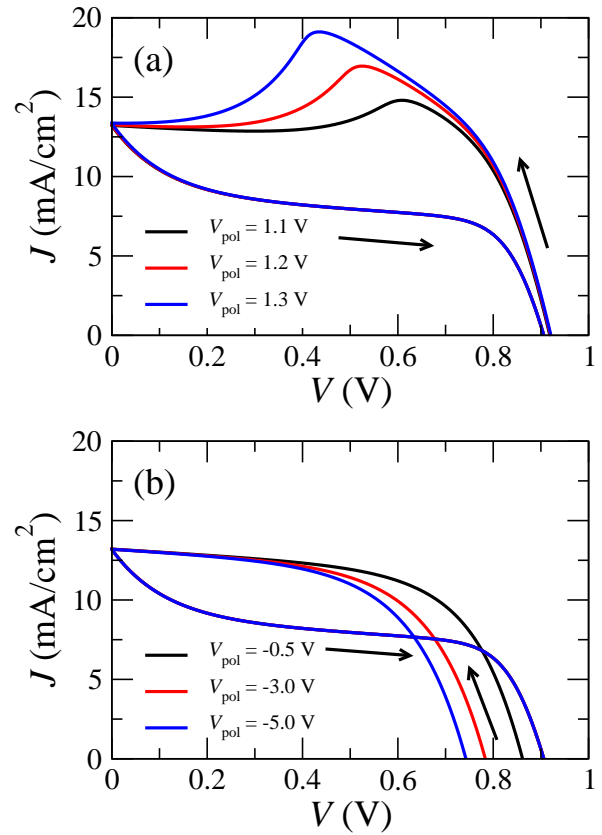}
\end{center}
	\caption{J-V characteristics for different poling conditions: (a) $V_{\rm pol}>V_{\rm oc}$ and (b) $V_{\rm pol}<0$, showing normal and mixed (inverted) hysteresis, respectively. The characteristic current bump is evidenced as the poling voltage is increased. Negative poling induces large recombination, visible in the reduced reverse current, due to the unfavorable electric field produced by the initial ionic displacement during the pre-poling phase. }
    \label{LS-Vpol}
\end{figure}

A typical J-V measurement in the dynamic regime is performed as a reverse-forward (R-F) scan, i.e. the applied voltage is varied from the initial value $V_0 \approx V_{\rm oc}$ to short-circuit and back. A proper evaluation and interpretation of the hysteretic effects is important for a correct assessment of the PCE \cite{NEMNES2018976}, but can also be regarded as a monitoring tool for degradation \cite{C8TC05999C}. Applying a poling voltage $V_{\rm pol} \gtrsim V_{\rm oc}$ produces normal hysteresis, while a negative value $V_{\rm pol} < 0$ results in a mixed or completely inverted hysteresis \cite{doi:10.1021/acs.jpcc.7b04248}. This behavior is shown in Fig.\ \ref{LS-Vpol} (a) and (b). Increasing $V_{\rm pol}$ in the sequence 1.2 V, 1.4 V and 1.6 V, the {\it current bump}, which is a hallmark of the reverse current scan, is getting larger. This is matched by J-V measurements under these pre-poling conditions and numerically reproduced by a few theoretical studies \cite{C5EE02740C,SolarEnergyMaterialsSolarCells.159.197.2017.Nemnes,doi:10.1021/acs.jpclett.7b00045}. Although both CA and CC models accurately describe the current bump, the physical microscopic picture is rather different: CA models predict large interfacial charges that can be accumulated during prepoling and which is then depleted during the reverse scan resulting in a large current in the reverse scan, while the CC models describe the same feature as an enhanced charge collection due to the properly oriented electric field created by the ion distribution. The modulation of the recombination current, ionic capacitance and ionic charge under poling conditions are detailed in Figs.\ S2(a), S3(a) and S4(a) in the SM, respectively. It is worth mentioning that the appearance of the bump requires a non-linear C-V dependence of the type outlined in Eq.\ (\ref{C_V}), while a constant capacitance would not reproduce the bump. On the other hand, increasing the absolute value of the negative poling voltage, in the sequence -1.5 V, -3 V, -5 V, drives the mixed hysteresis towards a better defined inverted hysteresis.

\begin{figure}[t]%
\begin{center}
  \includegraphics[width=0.98\linewidth]{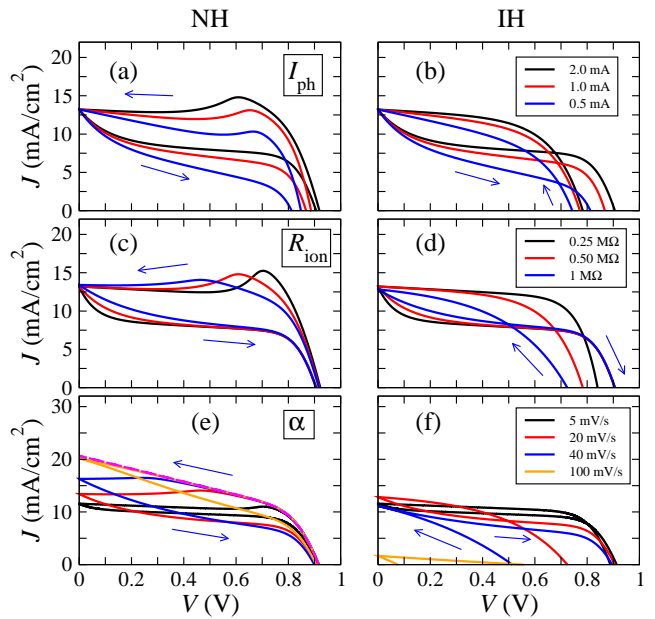}
\end{center}
	\caption{A comparative analysis of the dynamic J-V characteristics under normal- and inverted hysteresis conditions and varying: (a,b) illumination ($I_{\rm ph} = 0.5, 1, 2$ mA), where $J$ has been rescaled for the lowest two illumination intensities by 4$\times$ and 2$\times$ ; (c,d) ionic resistance ($R_{\rm ion}$ = 0.25, 0.5, 1 M$\Omega$); (e,f) scan rate ($\alpha=$ 5, 20, 40, 100 mV/s), for $R_{\rm ion} = 1$ M$\Omega$, where the limiting case of a very rapid scan ($1$V/s) is shown for NH (magenta).  The color codes are the same for NH and IH.}
    \label{LS-IphRcAlpha}
\end{figure}

Illumination is another important factor which influences the hysteretic effects. The measurements performed by Tress {\it et al.} show in Fig.\ 1(b) of Ref.\ \cite{C4EE03664F} a partial scaling of the J-V curves for a relatively wide range of illumination conditions. A similar picture is reproduced by our model, where the reference value $I_{\rm ph} = 2$ mA is reduced to 1 mA, then 0.5 mA, as depicted in Fig.\ \ref{LS-IphRcAlpha} (a) and (b). Near short-circuit the scaling is most evident. Two other features that seemingly depart from the perfect scaling are visible in the experimental study and found also in our simulations: (i) the shift of the current bump towards smaller voltages as the illumination is increased and (ii) the decrease of the $V_{\rm oc}$'s for either reverse or forward scans with decreasing illumination. These trends were also confirmed recently by dynamic J-V measurements \cite{doi:10.1021/acsenergylett.2c01252}.

The characteristic time scales observed in the hysteretic effects have a typical range of seconds to a few hundreds of seconds and they are correlated with the ionic migration processes. Essentially, the time scales associated with the slow processes are given by $\tau = R_{\rm ion}C_{\rm ion}$, where $C_{\rm ion}$ has an explicit voltage dependence. Keeping the scan rate constant and increasing the ionic resistance in the sequence $R_{\rm ion}$ = 0.25, 0.5, 1.0 M$\Omega$, a shift of the current bump is evidenced in the reverse characteristics for NH, while the IH becomes more pronounced, as described in Fig.\ \ref{LS-IphRcAlpha} (c) and (d). Conversely, keeping the PSC parameters fixed and modifying the bias scan rate $\alpha$ in the range of 5 mV/s to 1 V/s has a significant impact on both the magnitude of the hysteresis and the short-circuit current \cite{SolarEnergyMaterialsSolarCells.159.197.2017.Nemnes,doi:10.1021/acs.jpcc.7b04248,C8TC05999C,C4EE03664F}. The hysteresis magnitude is maximum for intermediate scan rates, for which the R-F scan time is comparable to the ionic relaxation time scale, as depicted in Fig.\ \ref{LS-IphRcAlpha} (e) and (f). On the other hand, it vanishes at either very slow scan rates (i.e. reaching the stationary case) or at very fast ones, when the slow process cannot follow the rapid voltage change. The behavior of the recombination current, ionic capacitance and ionic charge under different scan rates is illustrated in Figs.\ S2(b), S3(b) and S4(b) in the SM, respectively. Under NH conditions, the short-circuit current increases with $\alpha$, in contrast to IH. This kind of dynamic J-V measurements can reveal further information concerning ion mobilities and species types migrating in the perovskite absorber towards one or both interfaces. To this end, for a suitable characterization of the hysteresis magnitude, a proper definition for the hysteresis index was introduced which can account for both NH and IH in a balanced way \cite{NEMNES2018976}, using integral measures that reflect the dynamical process.

\subsection{Small-signal analysis}

The hysteresis presented in the dynamic J-V scans hints to underlying capacitive effects. The physical capacitance is identified as an ionic capacitance, while the apparent capacitance extracted by EIS is significantly larger under illumination and it is due to a modulation of the collected current or, equivalently, the recombination current, driven by the instantaneous electric field produced by the ion distribution.  

Not long after the reports of giant capacitive effects, a more peculiar inductive effect was observed in the EIS experiments. While the capacitive effects were subject to different interpretations, the inductive effects cannot be associated with a classical inductor and several explanations were proposed based on recombination or injection currents. In spite of these efforts there is currently no unified picture, while a potential relation between the capacitive and inductive effects is still missing.

\subsubsection{Capacitive effects}

\begin{figure}[t]%
\begin{center}
  \includegraphics[width=0.90\linewidth]{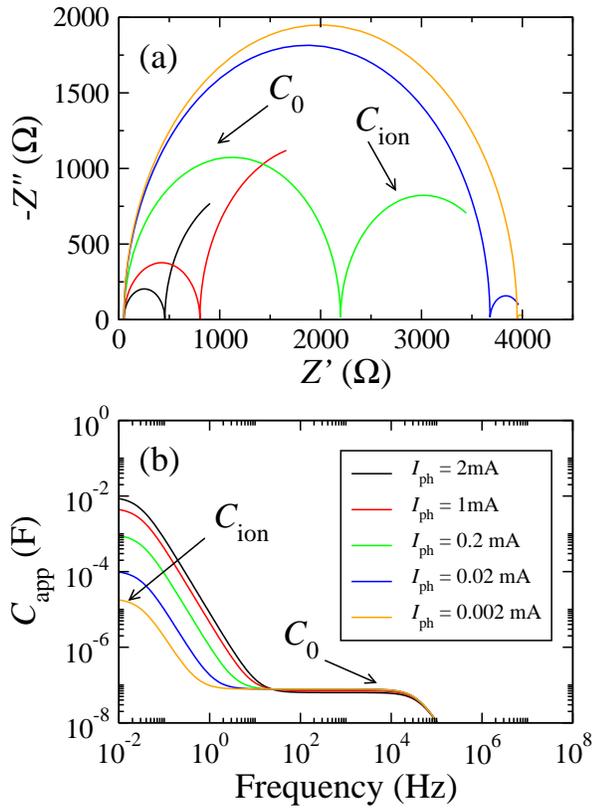}
\end{center}
	\caption{Capacitive effects evidenced by EIS analysis, for several illumination intensities, set by the photogenerated current $I_{\rm ph} = 2, 1, 0.2, 0.02, 0.002$ mA: (a) Nyquist plot and (b) the apparent capacitance ($C_{\rm app}$) as a function of frequency. Under practically dark conditions ($I_{\rm ph} = 0.002$ mA) the ionic capacitance ($C_{\rm ion}$) is found at low frequencies, while at higher frequencies the geometrical capacitance ($C_0$) is evidenced. However, $C_{\rm app}$ increases sharply with illumination intensity, as in typical EIS experiments.}
    \label{ss-iph}
\end{figure}

The significant enhancement of the apparent capacitance $C_{\rm app}$ by increasing the illumination is one key aspect, which can be understood by modulated recombination currents. We begin our simulated EIS analysis by considering in a first step only capacitive behavior without any inductive effects. Such a case is established by setting $a=0$ in Eq.\ (\ref{irec1cap}) and, consequently, the term $b\times I_{\rm c}$ controls the recombination current.  Figure\ \ref{ss-iph} shows the Nyquist plot analysis and the frequency dependence of $C_{\rm app}$, starting with practically dark conditions ($I_{\rm ph}=2\times 10^{-3}$ mA) and up to the maximum considered illumination intensity ($I_{\rm ph}=2$ mA). We consider a working point $V_{\rm wp}=0.7$ V, close the maximum PCE point, with the PSC in the stationary regime. Under dark conditions, two plateaus can be identified, one corresponding to the ionic capacitance ($C_{\rm ion} \simeq 10 \mu$F) and the other one corresponding to the geometrical capacitance ($C_0 \simeq 0.07 \mu$F). The data is consistent with the results shown in various reports of EIS measurements in similar conditions, as well as with our own experimental data, as it is shown in Section \ref{expval}. Correspondingly, the Nyquist plot shows two arcs, one belonging to the ionic capacitance at low frequencies and one to the geometrical capacitance at high frequencies.

\subsubsection{Inductive effects} 
\label{inductive_effects}

\begin{figure}[t]%
\begin{center}
  \includegraphics[width=0.90\linewidth]{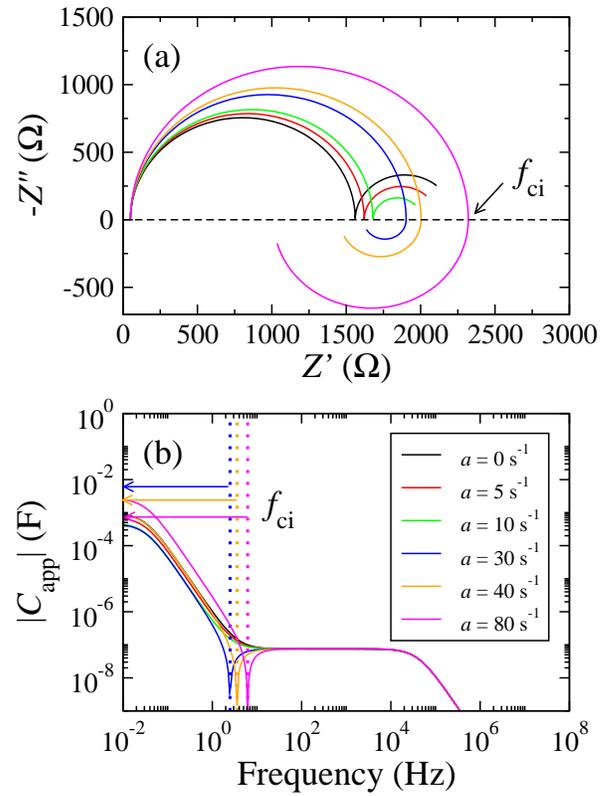}
\end{center}
	\caption{Inductive behavior produced by ion-defect induced recombination. Increasing the parameter $a$ in the sequence 0, 5, 10, 30, 40, 80 s$^{-1}$ the recombination current is enhanced leading to significant inductive effects. These are visible in the Nyquist plot (a) and they correspond to the frequencies ranges marked by arrows (b). The threshold frequency $f_{\rm ci}$ marks the transition between the capacitive and inductive behavior.}
\label{ss-a}
\end{figure}

Inductive effects may appear at small or intermediate frequencies in some PSCs. We argue here that the inductive behavior can be produced by defect induced recombination controlled by ionic accumulations at the interfaces, which is introduced by the term $a\times Q_{\rm C}$ in Eq.\ (\ref{irec1cap}). In Fig.\ \ref{ss-a} we assess the changes in the Nyquist plot and $C_{\rm app}-f$ dependence for $a = 0, 5, 10, 30, 40$ s$^{-1}$. Smaller values of $a$ still produce a capacitive behavior, while larger ones make the inductive effect visible at low frequencies, eventually overcoming the intrinsic capacitive behavior. To capture the sign change of Im$[Z]$, we considered in this case a lower value for the parameter that controls the capacitive behavior by the electric field induced recombination, $b=100$. 

Similar to the huge apparent capacitances observed under illumination, relatively large inductances can be observed in the EIS experiments, both having their origin in a modulated recombination current, which grows significantly with the illumination, as shown by experimental data in Section \ref{expval}. However, the inductive effects can be related to the ionic-induced defects, mostly located at the interfaces, as opposed to the capacitive effects that are due to the instantaneous electric field in the bulk of the perovskite. The ion mobility, as well as the susceptibility to induce defects is an important prerequisite for the inductive behavior. Therefore, the inductive effects may be regarded as a quantifiable feature for defect analysis and recovery in monitoring the PSC degradation.    

\begin{figure}[t]%
\begin{center}
  \includegraphics[width=0.98\linewidth]{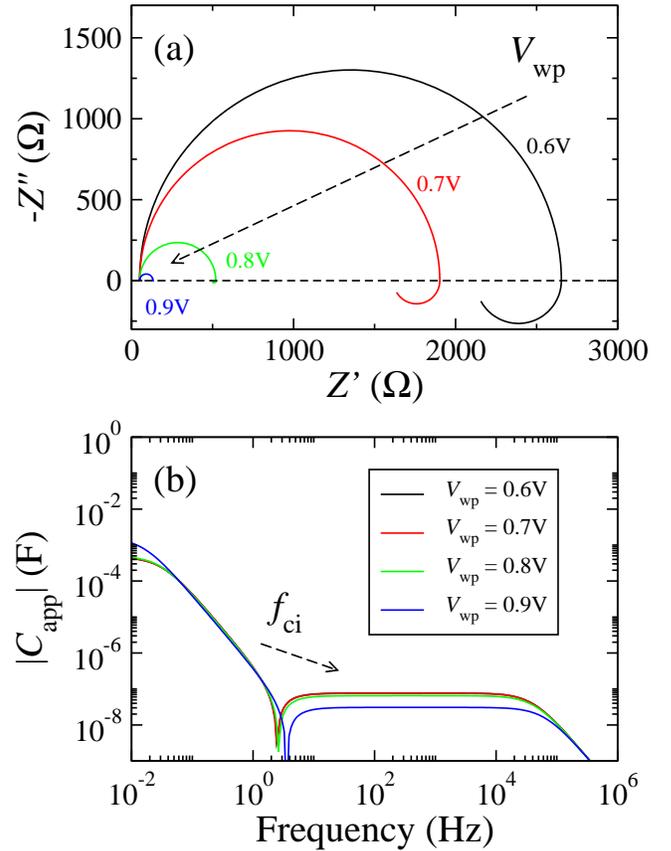}
\end{center}
	\caption{The influence of the working point ($V_{\rm wp}$) on the inductive behavior ($a = 30$ s$^{-1}$): (a) Nyquist plot showing an enhancement in the sample conductivity; (b) The absolute value of the apparent capacitance, indicating an increase of the threshold frequency ($f_{\rm ci}$) with $V_{\rm wp}$, where the capacitive effect is switched into an inductive effect and, also, a small increase in the apparent inductive effect by increasing $V_{\rm wp}$ at low frequencies.}
\label{ss-wp}
\end{figure}

The inductive behavior also depends on the working point pre-set in the EIS measurement, as it is shown in Fig.\ \ref{ss-wp}. Typically, the real part of the complex impedance decreases with $V_{\rm wp}$, signaling a more conductive behavior. The magnitude of the inductive effects increases at larger $V_{\rm wp}$ and low frequencies, due to larger ion accumulations. Furthermore, the transition point between the capacitive and inductive behavior, marked by $f_{\rm ci}$, drifts towards higher frequencies. These trends are in agreement with the experimental data, as shown in Section \ref{expval}.
Furthermore, the shift of $f_{\rm ci}$ with increasing $V_{\rm wp}$ induces an interesting conversion from a capacitor to an inductor as recently reported in Ref.\ \cite{doi:10.1021/acs.jpcc.2c02729}, which is illustrated in Fig.\ S6 in the SM. This may also support the idea that large ionic accumulations at positive voltages may induce the type of recombination that is responsible for inductive effects.

\begin{figure}[t]
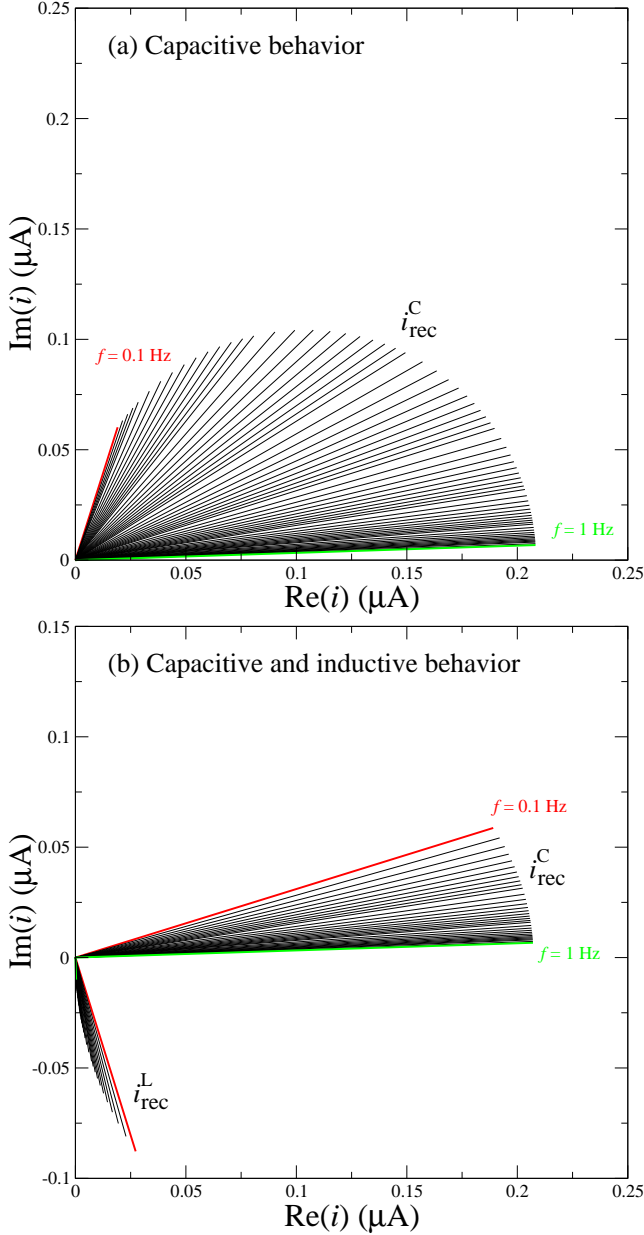
%
\begin{center}
\includegraphics[width=0.98\linewidth]{figure_9a}\\
\includegraphics[width=0.98\linewidth]{figure_9b}
\end{center}
	\caption{Phasor diagrams showing the dephasing of the recombination currents in the range of frequencies $f = 0.1 - 1$ Hz, for two types of PSCs, with: (a) capacitive effects only ($a=0$, $b=100$) and (b) capacitive and inductive effects ($a=30$ s$^{-1}$, $b=100$). The diagrams correspond to the data sets shown in Fig.\ \ref{ss-a}. The small signal voltage $v(t=0)$ has zero phase, i.e. it is oriented along the $x$-axis ($i=0$). In (b) subplot the $\pi/2$ dephasing between $i_{\rm rec}^{\rm C}$ and $i_{\rm rec}^{\rm L}$ is evidenced.}
\label{phasors}
\end{figure}

The capacitive and inductive behavior, as reflected by the two current components in Eq.\ (\ref{irec1cap}) applied for the small signal analysis, are suitably described by the phasor diagrams shown in Fig.\ \ref{phasors}. Applying a small signal with amplitude $v_0=1$ mV, the measured current $i$ is dephased by the two different recombination mechanisms. The parallel impedances $Z^{\rm eq}_{\rm L}$ and $Z^{\rm eq}_{\rm C}$ introduced in Eqs.\ (\ref{ZL}) and (\ref{ZC}) have the decisive contributions for this dephasing, where the small signal recombination currents $i_{\rm rec}^{\rm L}$ and $i_{\rm rec}^{\rm C}$ are found from $v+i R_s = i_{\rm rec}^{\rm L} Z^{\rm eq}_{\rm L} = i_{\rm rec}^{\rm C} Z^{\rm eq}_{\rm C}$. The currents $i_{\rm rec}^{\rm L}$ and $i_{\rm rec}^{\rm C}$ are dephased by $\pi/2$, as the former is proportional to $Q_c$ and the latter with ${\partial Q_c}/{\partial t}$. This can be easily checked in Fig.\ \ref{phasors}(b).
In the charge accumulation model, the impedance response would be equivalent to large capacitive ($C_{\rm acc} = 1.01$ mF) and inductive ($L = 16.6$ kH) elements, similar to previous reports \cite{doi:10.1021/acs.jpclett.7b00415}.

It is worth noting that there is a connection between capacitive and inductive effects, as the former depends on the ionic current, $\partial Q_{\rm c}/\partial t$, while the latter is connected to the ionic charge accumulations, $Q_{\rm c}$. The recombination current is, in general, a sum of these two components and the parameters $a$ and $b$, which are sample specific, will establish one of the two behaviors.

\begin{figure}[t]%
\begin{center}
\includegraphics[width=0.98\linewidth]{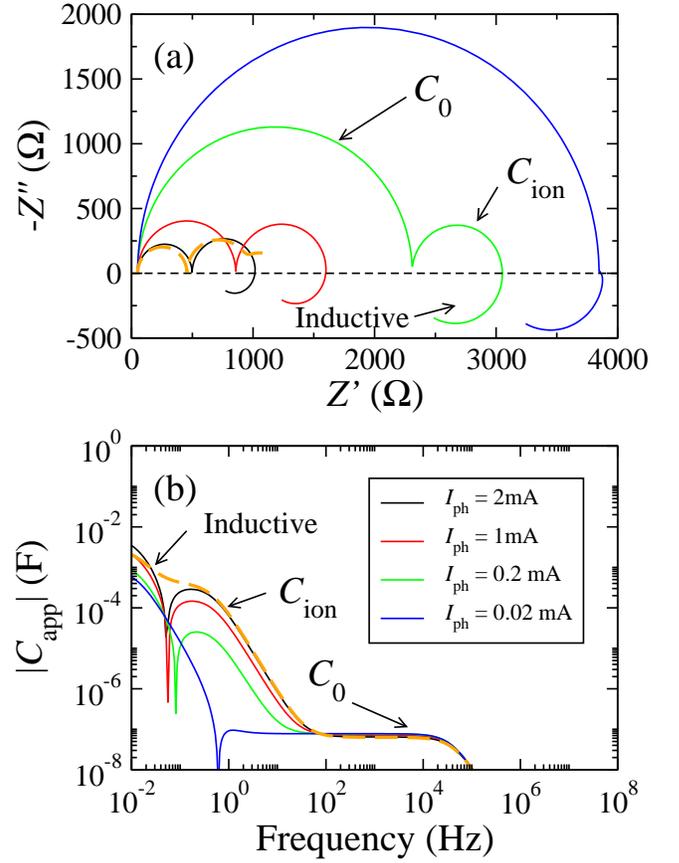}
\end{center}
	\caption{Two active interfaces with inductive effects: (a) Nyquist plot and (b) $C_{\rm app}-f$ plot. Three different illuminations ($I_{\rm ph} = 0.2, 1, 2$ mA) have been considered. The interfaces have different capacitances with the ratio $C_1/C_2=20$. Therefore, we identify the second semicircle $C_{\rm ion}$ to $C_2$ capacitance, while the arc corresponding to $C_1$ turns into the negative (inductive) region. The dashed lines correspond to vanishing inductive contribution for $I_{\rm ph} = 2$ mA.}
\label{two-interfaces}
\end{figure}

For some PSCs, two or three arcs in the Nyquist plot become visible \cite{TAN2021105658}, in addition to the low frequency inductive behavior, which is indicative of multiple relaxation time scales relevant for the capacitive behavior. The rather different time scales may originate from the migration of several ionic species (iodine, methylammonium ions, their charged vacancies) and/or their accumulation at both interfaces. The capacitances associated with the two interfaces can be rather different taking into account the ionic diffusivity at the ETL-perovskite and HTL-perovskite interfaces and the ion distribution.

Figure\ \ref{two-interfaces} describes the case with two ionic species (e.g. the negative iodine ions and the positive iodine vacancies) migrating towards both interfaces, with different time scales. The two capacitors are parametrized by $\bar{C}_{01} = 20$ $\mu$F, $\bar{C}_{11} = 2\times10^{-14}$ $\mu$F, $\bar{C}_{02} = 1$ $\mu$F, $\bar{C}_{12} = 10^{-15}$ $\mu$F, $n_c k_{\rm B}T = 26$ meV, i.e. we have $C_1/C_2=20$. In this case, the $Q_1$ is most significant and it generates the inductive effect at low frequencies, while $Q_2$ produces negligible inductive effects. The two capacitive arcs in the Nyquist plot correspond to $C_0$ and $C_2$. However, if $a_2>>a_1$, i.e. the second interface has a larger impact on the recombination current, intermediate-frequency current loops can develop, as illustrated in Fig.\ S7 in the SM. Further analysis regarding this aspect, as well as concerning electron/hole-ion recombinations introduced by $R_{1,2}$ resistances is presented in Fig.\ S8 in the SM.

\subsection{Experimental validation}
\label{expval}

\begin{figure}[t]%
\begin{center}
\includegraphics[width=0.98\linewidth]{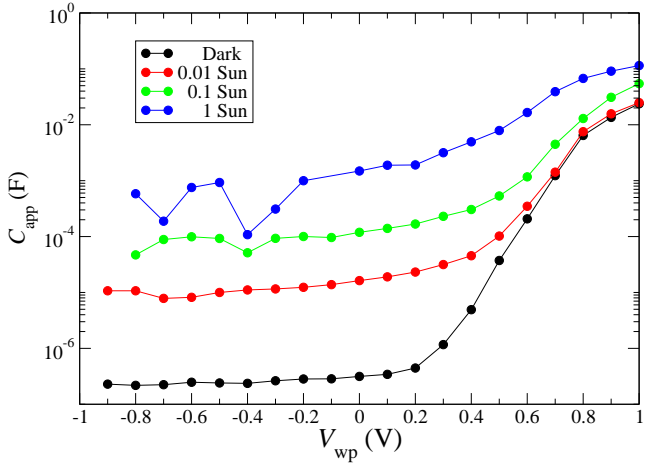}
\end{center}
	\caption{Experimental C-V data showing exponentially sharp increases at positive voltages. The same behavior is found for increasing the illumination intensity, while the capacitance plateau shifts towards higher values. The data was measured at a frequency $f=0.1$ Hz.}
\label{exp_C-V}
\end{figure}

\begin{figure}[t]%
\begin{center}
\includegraphics[width=0.98\linewidth]{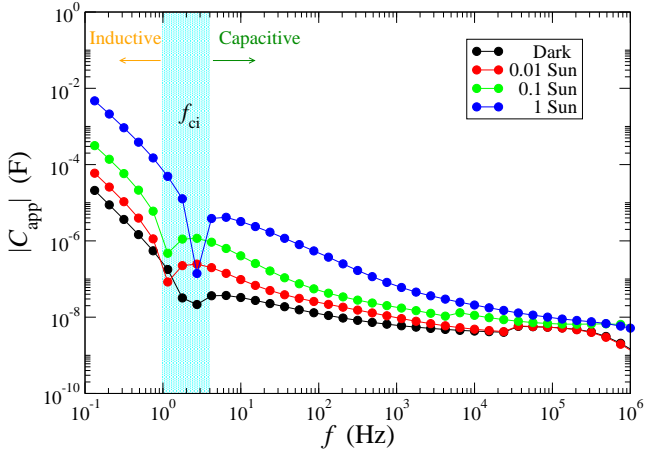}
\end{center}
	\caption{Measured C-f dependence obtained from the EIS analysis, for several illumination intensities, choosing the working point $V_{\rm wp} = 0.5$ V. Inductive behavior is found in the lower frequency range, with a significant increase with illumination intensity, indicating the role of the recombination currents of photogenerated carriers. The transition frequencies, $f_{\rm ci}$, are located in a narrow frequency interval, marked by the colored rectangle.}
\label{exp_C-f}
\end{figure}

\begin{figure}[t]%
\begin{center}
\includegraphics[width=0.98\linewidth]{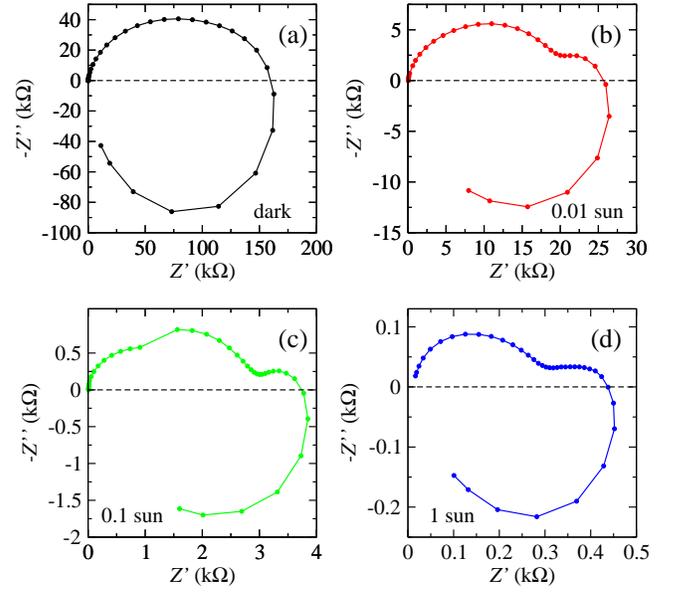}
\end{center}
	\caption{Nyquist plots obtained from EIS measurements, obtained for a working point $V_{\rm wp} = 0.5$ V, under different illumination conditions: (a) dark, (b) 0.01 sun, (c) 0.1 sun and (d) 1 sun. Two capacitive semi-circles and an inductive behavior at low frequencies are evidenced.}
\label{exp_Nyquist}
\end{figure}

\begin{figure}[t]%
\begin{center}
\includegraphics[width=0.98\linewidth]{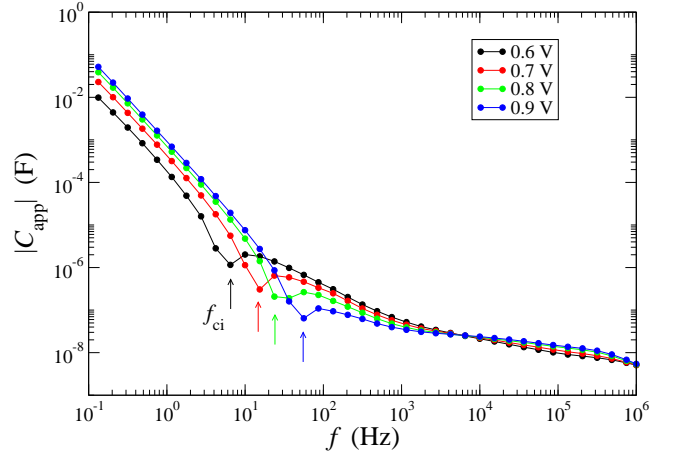}
\end{center}
	\caption{Experimental C-f data showing the shift of the transition frequency, $f_{\rm ci}$, with the applied working point, $V_{\rm wp} = 0.6, 0.7, 0.8, 0.9$ V, while there is also a small increase of the inductive effects in the low frequency range.}
\label{exp_C-f_voltage}
\end{figure}

While the large signal analysis based on dynamic J-V characteristics has been investigated in numerous studies and detailed features of the hysteric effects have been linked to specific measurement conditions \cite{NEMNES2018976}, the intriguing capacitive effects were highly debated, while the inductive behavior remains rather elusive. In the following, we present a selection of the most representative types of EIS measurements. 

The investigated PSC samples have a standard configuration consisting of planar layers as follows: glass substrate coated with fluorine-doped tin oxide (FTO), compact and mesoporous TiO2 layers deposited by spray coating, CH$_3$NH$_3$PbI$_{2.6}$Cl$_{0.4}$ mixed halide perovskite and spiro-OMeTAD deposited by spin coating and a sputtered Au top contact. Details about the performances of the investigated cells and preparation techniques are given in Refs. \cite{doi:10.1021/acs.jpclett.6b02375,https://doi.org/10.1002/ente.201900922}.

Electrochemical impedance spectroscopy (EIS) responses of the fabricated solar cells were measured with a broadband dielectric spectrometer model Alpha-A High Performance Frequency Analyzer from Novocontrol (Montabaur, Germany). The EIS experiments were performed in ambient laboratory conditions, using an alternating current (a.c.) potential of 10 mV amplitude, frequencies in the 10$^{-1}$ Hz $-$ 10$^7$ Hz range, d.c. potentials between -1 V and + 1 V and different illumination conditions given by a solar simulator, with the incident power varying between 0 and 100 mW/cm$^2$. A mask was used to ensure that only the active area of 0.09 cm$^2$ of the device was illuminated during the EIS measurements.

The Ansatz represented by Eq.\ (\ref{C_V}), introduced in Refs.\ \cite{SolarEnergyMaterialsSolarCells.159.197.2017.Nemnes,Anghel_2019}, concerning the exponential behavior of the capacitance as a function of applied voltage, in the dark or under illumination, was confirmed experimentally. The data shown in Fig.\ \ref{exp_C-V} was collected for a frequency $f = 100$ Hz and different illumination intensities, although it is worth mentioning that a similar behavior is found for a wide range of frequencies (1 Hz - 1 MHz range). This is in close connection with a well known feature in the J-V hysteresis under pre-poling conditions with $V>V_{\rm oc}$ -- the current bump in the reverse characteristics, which is a direct consequence of the sharp increase of the capacitance close to $V_{\rm oc}$. A mere constant capacitor would not produce this effect, although a rather flat J-V hysteresis would still be obtained for sufficiently large capacitances. 

Important insights can be drawn from analyzing the capacitive and the inductive behavior under illumination. Fig.\ \ref{exp_C-f} showing the apparent capacitance-frequency dependence is illustrative to this end. In dark conditions, the apparent capacitive and inductive effects (negative capacitances) reach moderate values ($10^{-5}-10^{-4}$ F), as a consequence of ionic response and limited recombination. As the illumination is increased up to 1 sun, significant apparent capacitances and inductances emerge, which emphasizes the role of the recombination of photogenerated carriers. For capacitive behaviors this light-induced enhancement is rather well documented \cite{doi:10.1021/acs.chemrev.1c00214}, in contrast to a similar trend for inductive effects under illumination. Typical Nyquist plots are shown in Fig.\ \ref{exp_Nyquist}, which indicates the impedance reduction with illumination.

Furthermore, Fig.\ \ref{exp_C-f_voltage} shows the modifications in the apparent capacitive/inductive behaviors as the working point, $V_{\rm wp}$, is changed from 0.6 to 0.9 V. The transition point ($f_{\rm ci}$) shifts to higher values as $V_{\rm wp}$ is increased, while the inductances are enhanced by approximately one order of magnitude. These observations are consistent with the measurements performed by Ebadi {\it et al.}, shown in Fig.\ 2(e) of Ref.\ \cite{Ebadi2019} and also with our simulations presented in Section\ \ref{inductive_effects}.

Overall, the EIS experimental data supports the idea of ion-induced recombination of photogenerated carriers, as the magnitude of both capacitive and inductive effects is in strict correlation with the illumination intensity.\\   

\section{Conclusions}

We introduce a comprehensive equivalent circuit model able to reproduce both capacitive and inductive effects. Starting with the large signal analysis, we recover known dynamical effects experimentally observed, such as the current bump in the reverse scan under positive poling, inverted hysteresis under negative poling and tuning of the hysteretic effects under illumination and bias scan rate, correlated with the relatively large time scales involved in the ion migration processes. 

Our analysis provides a bridging point between the charge accumulation 
models and charge collection models. 
Regarding the capacitive behavior, we show
that both models can lead to similar results, being formally
equivalent, provided the time scale of the slow process is the same.
An important difference is that the CA model uses a small resistance ($\sim 100 \ \Omega$), but a large capacitance ($\sim 0.1$ F) 
to match a typical time scale of $\sim 10$ s. The huge charge associated
with this capacitance is unlikely to exist in the device. Instead, in the CC
model, the same time scale is obtained with a much smaller ionic capacitance 
($\sim 100$ $\mu$F), but a larger resistance ($\sim 100 \ {\rm k}\Omega$)
associated to an ionic current. In this case, the recombination current depends on the 
intrinsic electric field of the device, which includes the contribution 
of the displaced ions. 

Concerning the inductive effects, the CA model uses
an explicit inductive element in the equivalent circuit, whereas  
in the CC model the inductance is associated to a dephased recombination current. 
This association is based on the Ansatz that one contribution to the 
recombination current comes from the ionic charge accumulations
($I_{\rm rec}^{\rm L}\sim Q_{\rm c})$.

The small signal analysis provides a clear identification of the contributions
to the recombination current, which are associated with capacitive
and inductive effects, respectively: the ionic current is linked to
the electric field in perovskite absorber, which, in turn, controls the
electron-hole recombination, while the ionic charge accumulations at the
interface determine the amount of defects modulating the recombination
current.

The experimental results confirm the assumption concerning the voltage dependence of the capacitance, which is essential for describing the J-V hysteresis and reinforce the connection between the capacitive and inductive effects, both being linked to the recombination of photogenerated carriers. The strong variation of the inductive effects with illumination is indicative for a mechanism based on the recombination of photogenerated carriers, emerging from ionic accumulations. This complements the other recombination mechanism based on local electric fields in the absorber, which is responsible for the capacitive effects. 

Furthermore, this investigation creates the framework for assessing the defects induced by ion migration, particularly by monitoring the inductive effects over time, which can become a predictive tool for the long term degradation of the PSC by external factors or by bias stress tests. \\

{\bf Acknowledgements}\\

The authors acknowledge the financial support of EEA Grants 2014-2021, under Project contract no. 36/2021.



\bibliography{manuscript_R2-C1}

\onecolumngrid




\setcounter{figure}{0}

\makeatletter 
\renewcommand{\thefigure}{S\@arabic\c@figure}
\makeatother

\section*{Supplemental Material}

%


\begin{flushleft}
	{\bf 1. Large-signal analysis}
\end{flushleft}

\begin{figure}[h]%
\begin{center}
\includegraphics[width=0.55\linewidth]{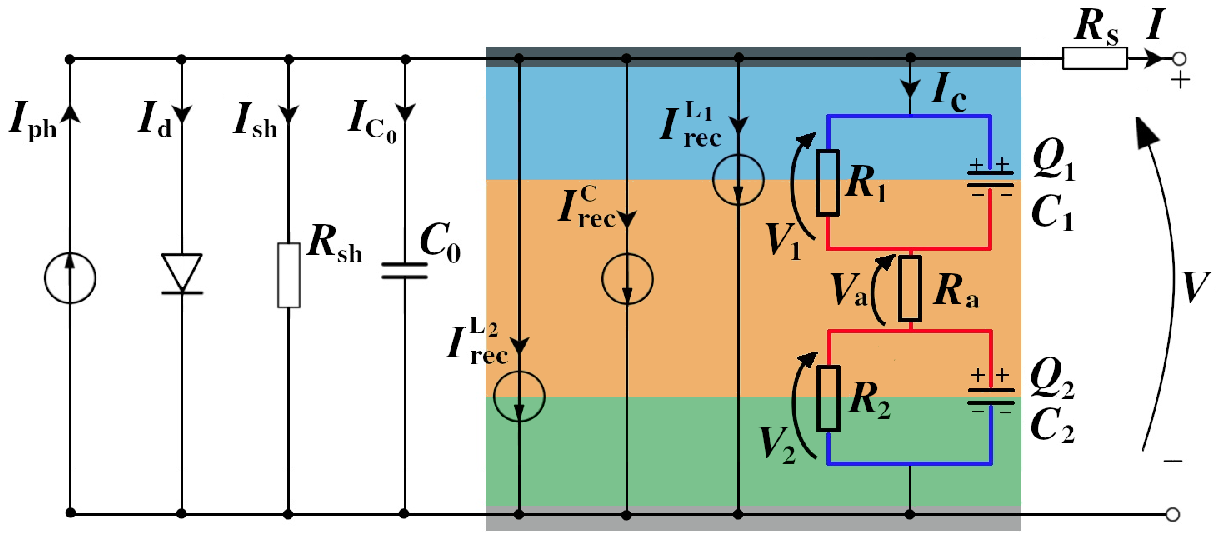}\hspace*{0.5cm}
\end{center}
	\caption{Large signal circuit model showing a detailed description of the ion-induced recombination current, which includes the inductive contributions ($I_{\rm rec}^{\rm L_1}=a_1Q_1$ and $I_{\rm rec}^{\rm L_2}=a_2Q_2$) associated with the ionic charge accumulations, $Q_{1,2}$, and a capacitive contribution ($I_{\rm rec}^{\rm C}=bI_{\rm c}$), which is due to the electric field in the bulk of the perovskite, $\vec{\mathcal E}$, and can be related to the ionic current, $\vec{J}_c = \sigma_{\rm ion} \vec{\mathcal E}$. In the circuit branch containing $C_1$ and $C_2$ capacitors, the pathways of ions and of electrons/holes are depicted in red and blue colors, respectively.} 
\label{DetailedLSC}
\end{figure}

\begin{figure}[h]
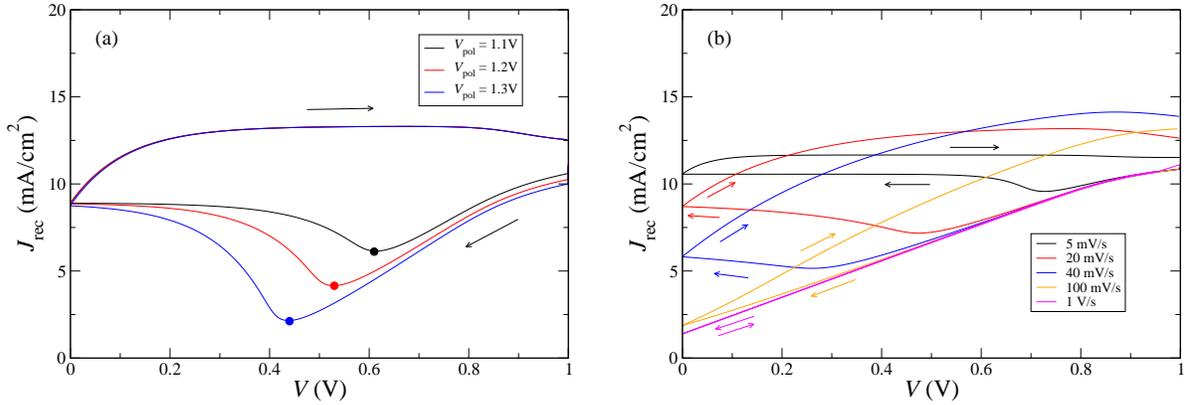
%
\begin{center}
\includegraphics[width=0.42\linewidth]{figure_SM_2a}\hspace*{0.5cm}
\includegraphics[width=0.42\linewidth]{figure_SM_2b}
\end{center}
	\caption{Current {\it bump} origin explained by the recombination current modulated by the electric field in the perovskite, under pre-poling and bias conditions (reverse-forward scan): (a) influence of pre-poling voltage ($V_{\rm pol} = 1.1, 1.2, 1.3$ V), with lowest recombination (highest carrier collection) at the bump location marked by colored dots and (b) influence of the bias scan rate ($\alpha = 5, 20, 40, 100, 1000$ mV/s). The data corresponds to the J-V characteristics shown in Figs. 4(a) and 5(e), respectively.}
\label{SI_Irec}
\end{figure}

\vspace*{3.5cm}

\begin{figure}[h]
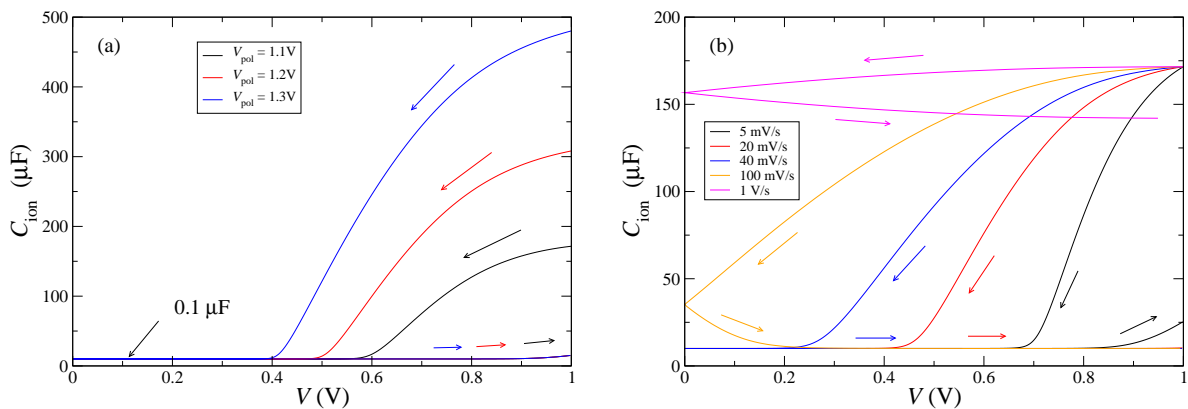
%
\begin{center}
\includegraphics[width=0.42\linewidth]{figure_SM_3a}\hspace*{0.5cm}
\includegraphics[width=0.42\linewidth]{figure_SM_3b}
\end{center}
	\caption{The ionic capacitance vs. applied voltage in the same conditions as in Fig.\ \ref{SI_Irec}, depending on: (a) pre-poling voltage and (b) bias scan rate.}
\label{SI_Cion}
\end{figure}

\vspace*{2.0cm}

\begin{figure}[h]
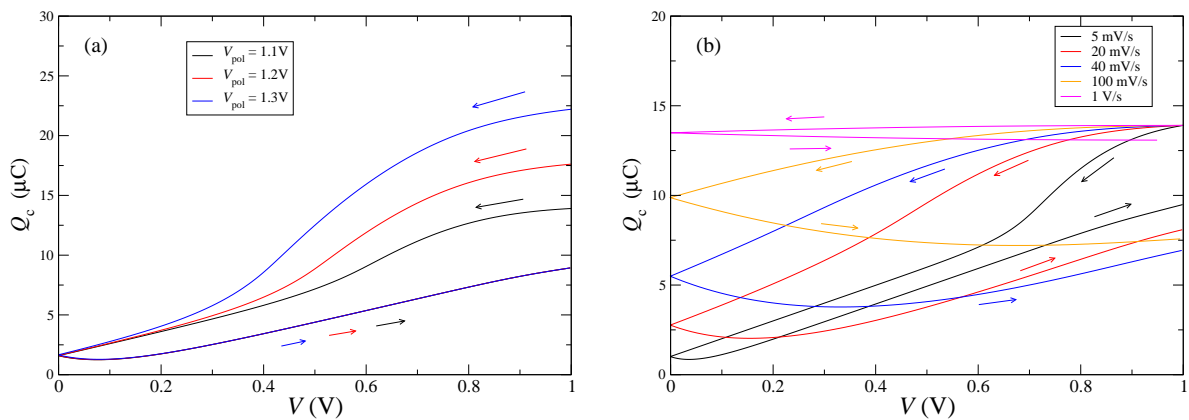
%
\begin{center}
\includegraphics[width=0.42\linewidth]{figure_SM_4a}\hspace*{0.5cm}
\includegraphics[width=0.42\linewidth]{figure_SM_4b}
\end{center}
	\caption{The ionic charge ($Q_c$) vs. applied voltage in the same conditions as in Fig.\ \ref{SI_Irec}, depending on: (a) pre-poling voltage and (b) bias scan rate.}
\label{SI_Qc}
\end{figure}

\newpage

\begin{flushleft}
	{\bf 2. Small-signal analysis}
\end{flushleft}

\begin{figure}[h]%
\begin{center}
(a) \includegraphics[width=0.6\linewidth]{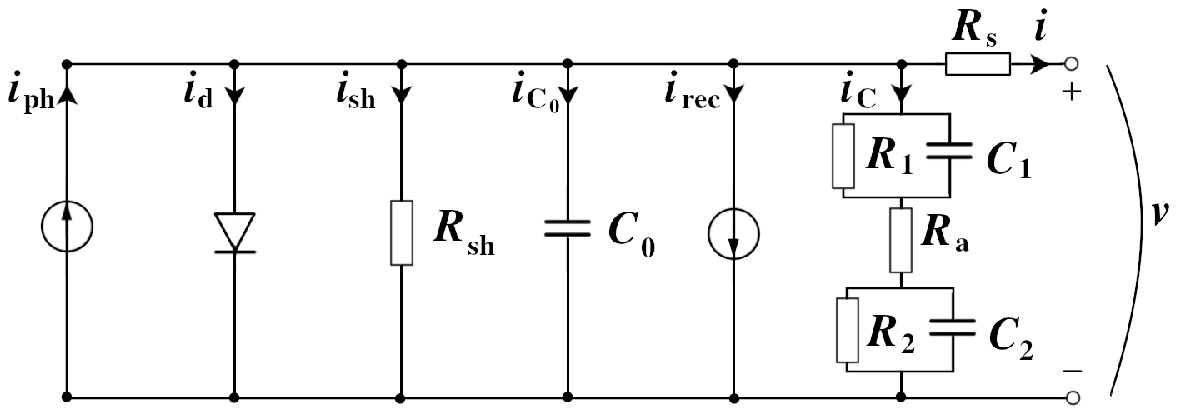} \vspace*{0.5cm}\\
(b) \includegraphics[width=0.6\linewidth]{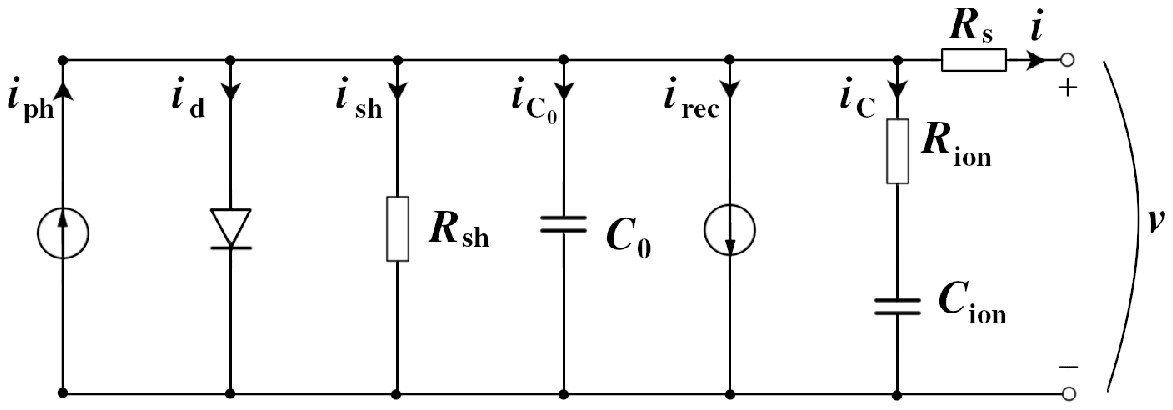}
\end{center}
	\caption{Equivalent circuits for small signal analysis: (a) Two-interface circuit model with two ionic capacitances ($C_1$ and $C_2$), ionic capacitors' loss resistances ($R_1$ and $R_2$) and absorber ionic resistance ($R_{\rm a }$), in correspondence with Fig.\ \ref{EQC_mod_3}; (b) Single interface circuit model, with one ionic capacitor ($C_{\rm ion}$) and the ionic resistance ($R_{\rm ion}$), in correspondence with Fig.\ \ref{CA-1cap}(b). This is obtained by setting: $R_1\rightarrow\infty$, $R_2\rightarrow0$, $C_1=C_{\rm ion}$ and $R_a = R_{\rm ion}$ in the more general circuit shown in Fig.\ \ref{SI_equivcirc}(a). } 
\label{SI_equivcirc}
\end{figure}

\begin{figure}[h]%
\begin{center}
\includegraphics[width=0.5\linewidth]{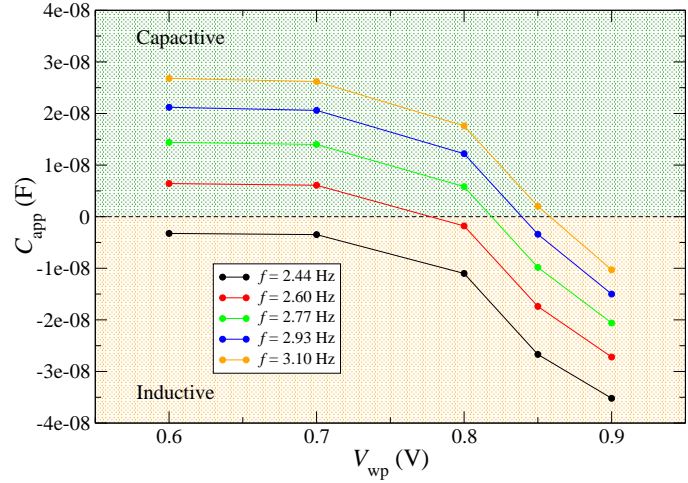}
\end{center}
	\caption{Changing behavior from capacitive to inductive by increasing the working point voltage. The sign change in $C_{\rm app}$ is associated with the observed shift in capacitive to inductive transition frequencies, $f_{\rm ci}$, in Fig.\ \ref{ss-wp}. At low frequencies, one can find an inductive behavior for the considered bias range, while at higher  frequencies the capacitive behavior is dominating. }
\label{CapInd}
\end{figure}

\vspace*{1.5cm}

\begin{figure}[h]%
\begin{center}
\includegraphics[width=0.95\linewidth]{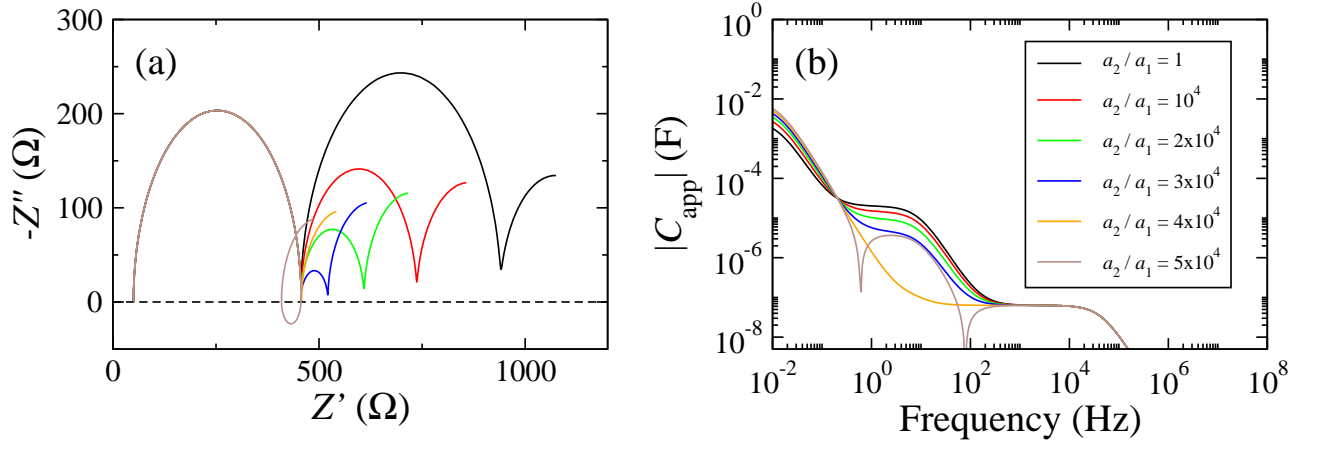}
\end{center}
	\caption{Inductive loops at intermediate frequencies: (a) Nyquist and (b) C-f plots. The two capacitors are parametrized by $\bar{C}_{01} = 20$ $\mu$F, $\bar{C}_{11} = 20\times10^{-15}$ $\mu$F, $\bar{C}_{02} = 0.05$ $\mu$F, $\bar{C}_{12} = 0.05\times10^{-15}$ $\mu$F, $n_c k_{\rm B}T = 26$ meV, i.e. we have $C_1/C_2=400$. We consider in this case $a_1=1$ and $a_2/a_1$ in the sequence specified in the legend. A smaller $C_2$ capacitance drives the inductive effects at higher (intermediate) frequencies, producing the inductive loop. As $Q_2$ is correspondingly smaller, the ion-induced recombination effects become visible as $a_2$ is increased.}
\label{f_a2}
\end{figure}

\vspace*{1.5cm}

\begin{figure}[h]%
\begin{center}
\includegraphics[width=0.95\linewidth]{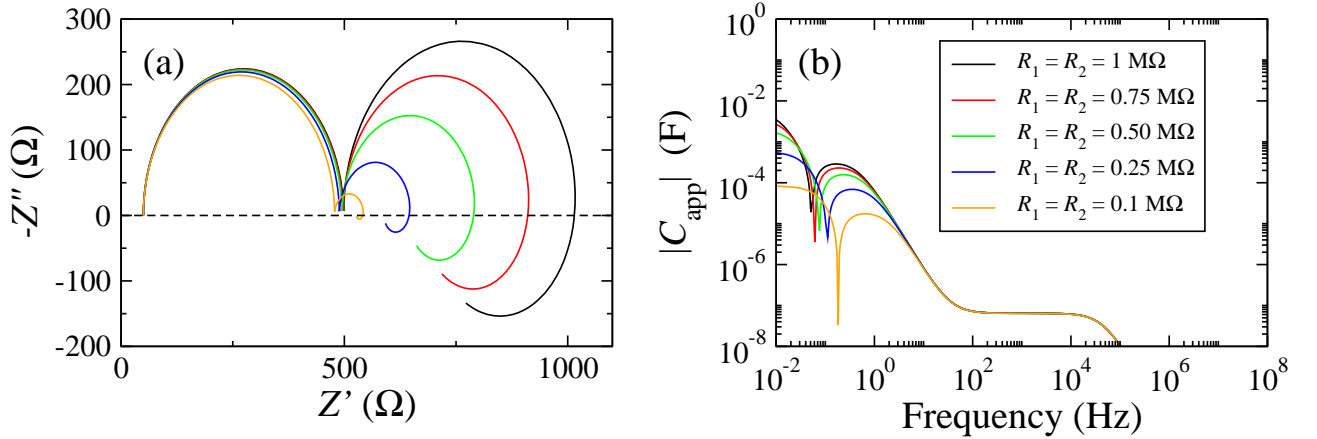}
\end{center}
	\caption{The influence of the ionic capacitors' loss resistances, $R_1$ and $R_2$: (a) Nyquist plots and (b) C-f dependence. The capacitive / inductive behaviors at low frequencies are reduced as $R_{1,2}$ become smaller (enhanced recombination). This trend becomes more visible as $R_{1,2} < R_{\rm a} = 0.5$ M$\Omega$.}
\label{R1_R2}
\end{figure}

\vspace*{3cm}

\newpage

 \begin{table}[!htbp] \centering
	 \caption{Description of small-signal circuit elements, corresponding to the circuits shown in Fig.\ \ref{SI_equivcirc}.}
 \label{Ztab}
 \begin{tabular}{@{\extracolsep{5pt}} cc} 
 \\[-1.8ex]\hline 
 \hline \\[-1.8ex] 
	   {\bf Small-signal element} & {\bf Description} \\
 \\[-1.8ex]\hline  \hline \\[-1.8ex] 
          1. Standard elements &   \\
 \hline	 
	  $R_{\rm d}$ & Diode dynamic resistance -- effective junction recombination \\
	  $R_{\rm sh}$ & Shunt resistance -- recombination due to pinholes defects. \\ 
	  $R_{\rm s}$ & Series resistance -- resistance of layers and contacts. \\
	  $C_0$ & Geometrical capacitance. \\
 \\[-1.8ex]\hline  \hline \\[-1.8ex] 
          2. Ionic circuit elements & \\
 \hline
	  $C_1$, $C_2$ & Ionic capacitors located at HTL-perovskite, ETL-perovskite interfaces.\\  
	               & $ C_i(V_{i}) = \bar{C}_{0i} + \bar{C}_{1i} \exp\left(\frac{q_{\rm e}V_{i}}{n_c k_{\rm B}T}\right)$,  $i=1,2$ . \\
	  $R_1$, $R_2$ & Loss resistances associated with ionic capacitors $C_1$ and $C_2$. \\
	  $R_{\rm a}$ & Absorber resistance (ionic resistance).\\
 \\[-1.8ex]\hline  \hline \\[-1.8ex]
	  3. Impedances and recombination currents & \\
		    -- two active interfaces [Fig.\ \ref{SI_equivcirc}(a)] & \\
 \hline	  
	  $Z_{C_0}$ & Impedance corresponding to the geometrical capacitance, \\
	            & $Z_{C_0} = \frac{1}{j \omega C_0}$ . \\
	  $Z_0$ & Impedance corresponding to the parallel group $R_{\rm d} \parallel R_{\rm sh} \parallel Z_{C_0}$, \\
	        & $Z_0 = \left[\frac{1}{R_{\rm d}} + \frac{1}{R_{\rm sh}} + \frac{1}{Z_{C_0}} \right]^{-1}$ . \\
	  $Z_{R_i C_i}$ & Impedance of the parallel $R_i$-$C_i$ circuit blocks, \\
	                & $Z_{R_i C_i} = R_i/(j\omega R_i C_i + 1)$,  $i=1,2$ . \\
	  $Z_C$ & Impedance corresponding to the series group formed by $Z_{R_1 C_1}$, $R_{\rm a}$ and $Z_{R_2 C_2}$, \\
	        & $Z_C = Z_{R_1 C_1} + R_{\rm a} + Z_{R_2 C_2}$ . \\
	  $i_{\rm rec}^{\rm L}$ & Inductive current component: 
	          $i_{\rm rec}^{\rm L} = \sum_{i=1}^2 a_i C_i(V_{i}) Z_{R_i C_i} \times i_C$ . \\
	  $i_{\rm rec}^{\rm C}$ & Capacitive current component: 
		          $i_{\rm rec}^{\rm C} = b \times i_C$ . \\
	  $i_{\rm rec}$ & Current source corresponding to the ion-modulated recombination current, \\
                        & $i_{\rm rec} = i_{\rm rec}^{\rm L} + i_{\rm rec}^{\rm C} = \left[ \sum_{i=1}^2 a_i C_i(V_{i}) Z_{R_i C_i} + b \right] \times i_C$ . \\
	  $Z^{\rm eq}_{\rm L}$ & Equivalent impedance for the inductive contributions (two active interfaces), \\
	                       & $Z^{\rm eq}_{\rm L} = \left[ \sum_{i=1}^{2} a_i C_i(V_{i}) \frac{Z_{R_i C_i}}{Z_C} \right]^{-1}$ . \\
          $Z^{\rm eq}_{\rm C}$ & Equivalent impedance for the capacitive contributions (two active interfaces), \\
	                       & $Z^{\rm eq}_{\rm C} = \frac{Z_C}{(b+1)}$ .\\
	  $Z$ & Equivalent impedance of the small-signal circuit, \\	      
		      & $Z = R_{\rm s} + \left[\frac{1}{Z_0} + \frac{1}{Z^{\rm eq}_{\rm L}} + \frac{1}{Z^{\rm eq}_{\rm C}} \right]^{-1} 
		           = R_{\rm s} + \left[\frac{1}{Z_0} + \sum_{i=1}^{2} a_i C_i(V_{i}) \frac{Z_{R_i C_i}}{Z_C} + \frac{(b+1)}{Z_C} \right]^{-1}$ .\\
 \\[-1.8ex]\hline  \hline \\[-1.8ex]
	  4. Impedances and recombination currents & \\
	            -- one active interface [Fig.\ \ref{SI_equivcirc}(b)] & \\
 \hline	  
	  $i_{\rm rec}^{\rm L}$ &  $i_{\rm rec}^{\rm L} = -j \frac{a}{\omega} \times i_C$ . \\
	  $i_{\rm rec}^{\rm C}$ &  $i_{\rm rec}^{\rm C} =  b \times i_C$ . \\
	  $i_{\rm rec}$ & $i_{\rm rec} = i_{\rm rec}^{\rm L} + i_{\rm rec}^{\rm C} = -j \frac{a}{\omega} \times i_C + b \times i_C$ . \\
	  $Z_{\rm RL}$ & Impedance corresponding to the inductive contribution (one active interface), \\
		      & $Z_{\rm L} = \frac{1}{a C_{\rm ion}} + j\; \omega \frac{R_{\rm ion}}{a}$, ($R_1\rightarrow\infty$, $R_2\rightarrow0$, $C_1=C_{\rm ion}$, $R_a=R_{\rm ion}$) .\\
          $Z_{\rm RC}$ & Impedance corresponding to the inductive contribution (one active interface), \\	
	              & $Z_{\rm C} = \frac{R_{\rm ion}}{(b+1)} - j \frac{1}{\omega C_{\rm ion} (b+1)}$, ($R_1\rightarrow\infty$, $R_2\rightarrow0$, $C_1=C_{\rm ion}$, $R_a=R_{\rm ion}$) .\\
          $Z$ & Equivalent impedance of the small-signal circuit, \\	      
		      & $Z = R_{\rm s} + \left[\frac{1}{Z_0} + \frac{1}{Z_{\rm RL}} + \frac{1}{Z_{\rm RC}} \right]^{-1}$ . \\
 \\[-1.8ex] \hline \hline	\\[-1.8ex] 
 \end{tabular}
 \end{table}

\end{document}